\documentclass{elsart}

\usepackage{cite,graphicx,amsmath,amsfonts,amssymb,mathrsfs}

\newcommand{\ie}{{\it i.e.}}

\newcommand{\eg}{{\it e.g.}}

\newcommand{\cf}{{\it cf.}}

\newcommand{\eq}{Eq.}
\newcommand{\eqs}{Eqs.}

\newcommand{\fig}{Fig.}
\newcommand{\figs}{Figs.}

\newcommand{\Ref}{Ref.}
\newcommand{\Refs}{Refs.}
\newcommand{\Sec}{Sec.~}
\newcommand{\Secs}{Secs.~}

\newtheorem{definition}{Definition}

\begin{document}

\begin{frontmatter}

\begin{flushright}
{\small TUM-HEP-417/01}
\end{flushright}

\title{Decays of supernova neutrinos}

\author[labela]{Manfred Lindner\thanksref{label0}},
\thanks[label0]{E-mail: lindner@physik.tu-muenchen.de}
\author[labela,labelb]{Tommy Ohlsson\thanksref{label1}},
\thanks[label1]{E-mail: tohlsson@physik.tu-muenchen.de, tommy@theophys.kth.se}
\author[labela]{Walter Winter\thanksref{label2}}
\thanks[label2]{E-mail: wwinter@physik.tu-muenchen.de}

\address[labela]{Institut f{\"u}r Theoretische Physik, Physik-Department,
Technische Universit{\"a}t M{\"u}nchen, James-Franck-Stra{\ss}e, 85748
Garching bei M{\"u}nchen, Germany}

\address[labelb]{Division of Mathematical Physics, Theoretical
Physics, Department of Physics, Royal Institute of Technology (KTH), 100
44 Stockholm, Sweden}

\date{\today}
         
\begin{abstract}
Supernova neutrinos could be well-suited for probing neutrino decay,
since decay may be observed even for very small decay rates or coupling
constants. We will introduce an effective operator framework for the
combined description of neutrino decay and neutrino oscillations for supernova
neutrinos, which can especially take into account two properties: One is the
radially symmetric neutrino flux, allowing a decay product to be re-directed
towards the observer even if the parent neutrino had a different original
direction of propagation. The other is decoherence because of the long
baselines for coherently produced neutrinos. We will demonstrate how
to use this effective theory to calculate the
time-dependent fluxes at the detector. In addition, we will show the
implications of a Majoron-like decay model. As a result, we will demonstrate
that for certain parameter values one may observe some effects which could
also mimic signals similar to the ones expected from supernova models, making
it in general harder to separate neutrino and supernova properties.
\end{abstract}

\end{frontmatter}

\newpage

\section{Introduction}

Neutrino decay \cite{Bahcall:1972my,Pakvasa:1972gz} has been considered as an
alternative to neutrino oscillations, especially for atmospheric
\cite{Acker:1992eh,Barger:1998xk,Barger:1999bg,Fogli:1999qt,Lipari:1999vh,Choubey:1999ir}
and solar \cite{Acker:1992eh,Choubey:2000an,Bandyopadhyay:2001ct} neutrinos. 
Furthermore, sequential combinations of neutrino decay and
neutrino oscillations have been studied (\eg, MSW-mediated solar
neutrino decay \cite{Raghavan:1988ue,Bandyopadhyay:2001ct}), as well
as a combined treatment \cite{Lindner:2001fx}. Currently, one of the most
favorable models for neutrino decay is to introduce an effective
decay Lagrangian which couples the neutrino fields to a massless boson
carrying lepton number, \ie, a complex scalar field or a Majoron field
\cite{Zatsepin:1978iy,Chikashige:1980qk,Gelmini:1981re,Pakvasa:1999ta}. One
possibility for such a Lagrangian for the case of Majoron decay is
\begin{equation} \label{int2} \mathcal{L}_{\rm int} = \underset{i
\neq j}{\sum_{i} \sum_{j}} g_{ij} \overline{\nu_{j,L}^{c}} \nu_{i,L} J,
\end{equation}
where the $\nu$'s are Majorana mass eigenfields and $J$ is a Majoron field.
An interaction Lagrangian like this in general also implies coupling to
active neutrino mass eigenstates.
It was shown in \Ref~\cite{Lindner:2001fx} that interference effects,
such as neutrino oscillations, are in principle possible before and
after decay.
There are several constraints to Majoron models, such as the
exclusion of pure triplet models by the $Z^0$ width from LEP
at CERN. However, fine-tuning or small triplet admixtures to
singlet Majorons, as well as more sophisticated models have been
proposed to circumvent this problem. For a more detailed discussion
see, for example, \Refs~\cite{Acker:1992eh,Barger:1998xk,Barger:1999bg}.
Alternative neutrino decay models, such as the one mentioned in
\Ref~\cite{Schechter:1982cv}, proposing decay into more than one 
neutrino, have also been suggested. Nevertheless, since we are
interested in fast neutrino decay possibilities, we will focus on
models for decay of one neutrino into exactly one neutrino, which can
be described by a Lagrangian similar to \eq~(\ref{int2}).

In this paper, we extend the effective operator formalism introduced in \Ref
\cite{Lindner:2001fx} to allow a time-dependent treatment of
a supernova neutrino flux modified by neutrino decay and neutrino
oscillations. In earlier works
\cite{Frieman:1988as,Aharonov:1988ju,Soares:1989qt,Acker:1990tm},
different aspects of supernova decay have been considered, which we want to
combine into a general formalism.
For the sake of the analytical accessibility, however, we have to
make certain simplifications. Since we want to focus on the
neutrino propagation in vacuum, we thus ignore matter effects as well as
any other effect within the supernova. Nevertheless, as we will see, taking a
superposition of mass eigenstates obtained from a numerical
calculation within a supernova instead of an initial flavor state,
can be done in a quite straightforward way in our formalism.
Furthermore, we restrict ourselves to maximal one intermediate decay between
production and detection\footnote{Repeated decays correspond to higher order
processes suppressed by the small coupling constants. However, depending on
the decay model, for the very large supernova distances they may be relevant
making the geometrical discussion much more complicated. For the
implementation of repeated decays see also \Ref \cite{Lindner:2001fx}.} or
almost stable decay products\footnote{Since we assume that particles other
than the neutrinos, such as Majorons, are not detectable, the terms {\em decay
products} and {\em secondary neutrinos} will refer to the neutrinos produced
by decay. Note that in this paper the {\em secondary neutrinos} can be either
active or sterile neutrinos.}. We are especially interested in possible
modifications of the time dependence of a supernova neutrino signal.
Therefore, we neglect other effects such as (cosmological and
non-cosmological) redshift\footnote{We ignore redshift, since we know from
supernova statistics that the observable supernovas are quite close to the
Earth. In addition, we assume the supernovas not to be moving too fast within
the cosmological comoving frame.}, asymmetries of the supernova flux, and
spatial extension of the neutrino production region\footnote{Compared with the
distance to the Earth, supernovas can for our purpose be treated as point
sources. In addition, oscillation phases of neutrinos produced by a
core-collapse supernova are not necessarily averaged out by the extension of
the production region [$\mathcal{O}(10 \, \mathrm{km})$], which is typically
much smaller than the oscillation length in vacuum [$\mathcal{O}(1000 \,
\mathrm{km})$].}. However, we take into account dispersion by different
traveling times of different mass eigenstates, since we want to study the
physical effects of combining neutrino decay and neutrino oscillations. In
addition, we incorporate that neutrino oscillations are affected by wave
packet decoherence at long traveling distances if the neutrinos are coherently
produced or leaving the supernova as superpositions of mass eigenstates at
all. Since coherence of supernova neutrinos is still a
question under discussion, we will {\it a priori} assume that
supernova neutrinos are produced coherently as flavor states, but
demonstrate how to obtain the incoherent limit thereafter. When
taking into account the MSW effect, coherent emission can be achieved for
appropriate parameter values violating the adiabaticity condition. In
addition, for a three flavor neutrino scenario obtaining exactly one
(pure) mass eigenstate emerging from the supernova is no longer valid. It
may even be a superposition of two mass eigenstates, such that neutrino
oscillations of supernova neutrinos (close to the supernova) cannot be
excluded in general. Nevertheless, we will use the incoherent limit in most of
our examples, since this is the most often assumed case.

We will cover three main topics in this paper:
First, in \Sec \ref{Sec:decoh} we will discuss the loss of coherence
because of the long baselines and how this is modified by intermediate
decays. Especially, we will use Majoron decay as an example for demonstration.
Second, in \Sec \ref{Sec:framework}, we will introduce a general formalism in
order to be able to calculate time-dependent fluxes at the detector. It
will also implement the coherence issues mentioned before. Since it will
also work for decay models similar to Majoron decay, it will be
treated somewhat decoupled from the initial discussion. Third, in
the remaining sections, we will give some applications. In \Sec
\ref{Sec:massdisp}, we will demonstrate the time smearing of a
source pulse by neutrino decay. Coming back
to Majoron decay, in \Sec \ref{Sec:Majoron}, we will take a closer look on the
dynamics of a Majoron decay model and its implications. In the next section,
\Sec \ref{Sec:general}, we will calculate the flux at the detector by also
taking into account path lengths of different traveling paths. In addition, we
will apply the results for Majoron decay we obtained in \Sec \ref{Sec:Majoron}
in order to show the implications for this decay model. Finally, in \Sec
\ref{Sec:earlydecays}, we will show the peculiarity of observing interference
effects at the detector in the limit of quite large decay rates, even if the
coherence lengths of the problem are much shorter than the baseline length. In
almost all sections, we will use simple examples to visualize the indicated
effects.

\section{Coherence in neutrino oscillations and intermediate decays}
\label{Sec:decoh}

The coherence length for neutrino oscillations is of the order
$L^{\mathrm{coh}} \simeq {\sigma_x E^2 \over | \Delta m^2 |}$, where
$\sigma_x$ is the width of a (Gaussian) wave packet
\cite{Giunti:1991ca,Giunti:1998wq,Grimus:1998uh,Ohlsson:2000mj} or the wave
packet overlap in the source or detector \cite{Cardall:1999ze}. Thus, for
typical values for supernova neutrinos coherence seems to be destroyed on
their way to the Earth.
However, since the decay may
happen close to the source or detector, the distance between
decay and production or detection may be much shorter
than the coherence length. We will show that we may thus see coherent
interference effects in certain cases, \eg, for very early decays.
For this, we need to take into account that the wave packet width, which enters
the coherence length, is in fact
$\sigma_x^2=(\sigma_{x}^P)^2+(\sigma_{x}^D)^2$, \ie, the squared sum of the
widths of the production $P$ and detection $D$ processes \cite{Giunti:1998wq,Cardall:1999ze}. One can visualize this by a detection process working on a
much longer timescale than the production process. Therefore, though the wave
packets may be well separated to a third observer at arrival, the
detector will not be able to resolve them due to the time resolution of its
detection process. However, neutrino oscillations may be washed out, \eg, for
too large $\sigma_{x}^D > L^{\mathrm{osc}}$, since the detector then averages
over the oscillation length. For supernova neutrinos, the width is because of
$\sigma_x^P < 10^{-11} \, \mathrm{m} \ll \sigma_x^D$ determined by the
detection process only. Taking into account intermediate neutrino decays, the
detection process $D$ corresponds to the decay process $X$ with its width
$\sigma_x^X$. For Majoron decay of a relativistic mass eigenstate $i$ of
energy $E$ into a mass eigenstate $j$, determined by the Majoron coupling
constant $g_{ij}$, one can estimate the wave packet overlap by the decay rate
$\Gamma_{ij}$ as \begin{equation}   \label{SigmaM}  
\sigma_x^X \simeq {1 \over \Gamma_{ij}}  =
\mathcal{O} \left( \frac{E}{m_i m_j g_{ij}^2} \right),
\end{equation}
where the exact value of $\Gamma_{ij}$ depends on the interaction Lagrangian
\cite{Kim:1990km}. Thus, the weaker the coupling constants $g_{ij}$ are, the
slower is the process and the longer is the spatial extension of the wave
packet. Especially, for $g_{ij} \rightarrow 0$ the wave packet width would
become infinite. Nevertheless, we will see that neutrino oscillations are not
possible in this limit because of another constraint to be satisfied. Equation
(\ref{SigmaM}) leads to a coherence length for supernova neutrinos
\begin{equation}  \label{Lcoh}  L^{\mathrm{coh},X}_{ab} \equiv \frac{4
\sqrt{2} \sigma_x^X E^2}{\Delta m_{ab}^2} = \mathcal{O} \left( \frac{E^3}{m_i
m_j \Delta m_{ab}^2 g_{ij}^2} \right). \end{equation}
Here $\Delta m_{ab}^2$ is the mass squared difference of the
neutrino oscillation considered (before or after decay). For instance, an
incoming superposition of two mass eigenstates, oscillating in the flavor
basis, say $\nu_2$ and $\nu_3$, may decay into an outgoing
mass eigenstate, say $\nu_1$, by non-zero
Majoron coupling constants $g_{31}$ and $g_{21}$. Then the index $i$ of the
parent neutrino is $3$ or $2$, the index $j$ of the decay product is $1$, and
the $\Delta m_{ab}^2$ considered is equal to $\Delta m_{32}^2$.
{}From the last equation, \eq~(\ref{Lcoh}), we see that especially for small
coupling constants the coherence length can be quite long. However, in \Refs
\cite{Giunti:1998wq,Cardall:1999ze}, the upper bound
\begin{equation}
 L^{\mathrm{coh}}_{ab} < \frac{16 \pi^2 E^3}{(\Delta m_{ab}^2)^2 }
 \label{Lcohmax}
\end{equation}
was found for the coherence length, determined by the condition $\sigma_x <
L^{\mathrm{osc}}$ necessary for neutrino oscillations not to be washed
out by the spatial wave packet extension. Thus, this implies that
$L^{\mathrm{ coh}} < (10^9 \sim 10^{19}) \, \mathrm{m}$ for $\Delta m^2 \simeq
(10^{-5} \sim 1) \, \mathrm{eV}^2$, which is, in principle,  much shorter than
the typical distance of a supernova. In
addition, in order to observe the oscillations before or after decay, this
means for Majoron decay that
\begin{equation}  g_{ij}^2  >  \mathcal{O} \left(
\frac{\Delta m_{ab}^2}{m_i m_j} \right) \quad  \mathrm{or} \quad \alpha_{ij}
\equiv \Gamma_{ij} E  >  \Delta m_{ab}^2 = \mathcal{O} \left( \frac{\Delta
m_{ab}^2}{\mathrm{eV^2}} \, \frac{\mathrm{GeV}}{\mathrm{km}} \right).
\label{ConstraintM}
\end{equation}
This is a
finite lifetime condition similar to the one in \Ref \cite{Grimus:1998uh}
for the decaying muons used for neutrino production. 

Since we want to obtain a treatment which is independent of the
coherence discussion, we use the wave packet
approach in \Ref \cite{Giunti:1998wq} by assuming that coherence is destroyed
by a factor $\exp[ - (l/L^{\mathrm{coh}})^2]$. Here $L^{\mathrm{coh}}$ is
given by \eq~(\ref{Lcoh}) with the respective widths of the processes
considered. In addition, model-dependent constraints may have to be applied to
$L^{\mathrm{coh}}$. The following two cases will be discussed in this paper:
\begin{enumerate}
\item  Coherence is lost before decay and detection,
\ie, all coherence lengths involved in the problem are much shorter than the
traveling distances between the interaction processes. In the models
above, this can be easily achieved by violating \eq~(\ref{ConstraintM}). 
\item  All instable neutrinos have decayed before coherence is lost,
which means that
$\alpha L^{\mathrm{coh}}/E \gg 1$. Since for Majoron decay $\alpha$ as well as
$L^{\mathrm{coh}}$ depends on the coupling constants, one can show that this
condition together with \eq~(\ref{SigmaM}) implies that $E^2/\Delta m_{ab}^2
\gg 1$ in the wave packet model above applied to Majoron decay. This is always
true for relativistic neutrinos and small $\Delta m^2$'s, such as often assumed
for active neutrinos. In addition, constraints such as \eq~(\ref{ConstraintM})
may have to be satisfied for the neutrino oscillations considered.
\end{enumerate}
Note that \Secs \ref{Sec:massdisp} and \ref{Sec:general} are
calculated for the first case. The last
application, \Sec \ref{Sec:earlydecays}, corresponds to the second case.

\section{The formalism}
\label{Sec:framework}

In this section, we will introduce the formalism used to
calculate transition probabilities and fluxes. First, we will
motivate the extension of the formalism in \Ref \cite{Lindner:2001fx} by wave
packet aspects and properties of point sources. Then, we will define the
relevant operators, transition amplitudes, and fluxes. Finally, we will give
certain limiting cases in order to be able to describe realistic scenarios by
simplified formulas and to show that the formalism reduces to the one in \Ref
\cite{Lindner:2001fx} in the coherent limit.

\subsection{Motivation}

In \Ref \cite{Lindner:2001fx}, the combination of neutrino decay and neutrino
oscillations was discussed for beams, but not for point sources. Baselines
were assumed to be short enough such that the wave packets are still
sufficiently overlapping at the position of decay and at the
detector. Let us now generalize this with respect to the two above mentioned
aspects.

\subsubsection{Coherence and wave packets}
\label{Subsec:coherence}

In order to implement the loss of coherence by propagation over large
distances, $L>L^{\mathrm{coh}}$, we follow the mentioned wave packet treatment
of neutrino oscillations in \Refs \cite{Giunti:1991ca,Giunti:1998wq}. Therein
it was noted that using wave packets leads to additional factors in the
neutrino oscillation probabilities. The loss of coherence due to the spread and
different mean velocities of sharply peaked wave packets is described by a
factor $\exp [ - ( l/L^{\mathrm{coh},I}_{ab} )^2  ]$ in the
transition probabilities, where $L^{\mathrm{coh},I}_{ab}$ was defined in
\eq~(\ref{Lcoh}) and $\sigma_x^I$ is the spatial width of the wave packet
determined by the production $P$ and detection $D$ processes $(\sigma_x^I)^2 =
(\sigma_x^P)^2+ (\sigma_{x}^D)^2$. In addition, a factor enters the
neutrino oscillation formulas, which is equal to unity if the constraint in
\eq~(\ref{Lcohmax}) holds \cite{Giunti:1998wq}. Violating this constraint
formally also can be achieved by making the coherence length very
short.

Taking into account an intermediate decay process $X$ between production and
detection, which corresponds to the detection process above, we expect a factor
$\exp [ - ( l_1/L^{\mathrm{coh},I}_{ab} )^2  ]$ in the transition
probabilities, where $(\sigma_x^I)^2 = (\sigma_x^P)^2+ (\sigma_{x}^X)^2 \simeq
(\sigma_{x}^X)^2$ is the wave packet width for supernova neutrinos, $l_1$ is
the distance from production to decay, and the indices $a$ and $b$ refer to
the mass eigenstates before decay. This factor describes the loss of
coherence in the {\em in} state of a decay process, \ie, interference effects
caused by fixed relative phases in an incoming superposition of mass
eigenstates are destroyed. Furthermore, in decay new wave packets are
produced. This leads to additional factors $\exp [ - (
l_2/L^{\mathrm{coh},J}_{cd} )^2  ]$ describing the coherence among the decay
products $c$ and $d$ over the distance $l_2$ from decay to detection. In this
case, the wave packet width $(\sigma_x^J)^2 =(\sigma_{x}^X)^2 +
(\sigma_{x}^D)^2$ is determined by the widths of the decay and detection
processes.  In \Sec \ref{Subsec:operators}, we will introduce an operator
$\mathcal{S}$, which has the suggested properties.

\subsubsection{Radially symmetric source fluxes and time dependence}
\label{Subsec:time}

The second aspect we want to integrate into our operator framework is
time-dependent, radially symmetric source fluxes. We define
$\Phi^{\mathrm{tot}}_{\alpha}(t) = dN_{\alpha}/dt$ to be the total {\em source}
flux, \ie, the number of neutrinos of the flavor $\nu_{\alpha}$ emitted
equally in any direction per time unit. Hence, $\int_{-\infty}^{\infty}
\Phi^{\mathrm{tot}}_{\alpha}(t) dt =N_{\alpha}$ is the total number of
neutrinos emitted by the source. In addition, let $\Phi^{D}_{\alpha \beta}(t)$
be the ``flux'' at the detector, \ie, the number of neutrinos 
per time unit produced as flavor $\nu_{\alpha}$ and detectable
as flavor $\nu_{\beta}$. Since this flux knows about the produced
flavor, it could also be calculated by the transition probability as well as
the number of $\nu_{\beta}$ arriving per time unit at the
detector, \ie, $\Phi^D_{\beta}$. Often the term {\em flux} is used for
the function $\Phi^D_{\beta}$. However, in our notation
$\Phi^D_{\beta} = \sum_{\alpha} \Phi^{D}_{\alpha \beta}$. Furthermore, we will
split $\Phi^{D}_{\alpha \beta}(t)$ into $\Phi^{D,0}_{\alpha \beta}(t)$ and
$\Phi^{D,1}_{\alpha \beta}(t)$, describing neutrino fluxes with no
intermediate decays and one intermediate decay such as the transition
probabilities $P_{\alpha \beta}^0$ and $P_{\alpha \beta}^1$ in \Ref
\cite{Lindner:2001fx}. These may either have to be added or not, depending on
if the detector can distinguish undecayed particles from decay products (such
as by energy resolution or spin) or not.

\begin{figure}[ht!] \begin{center}
\includegraphics*{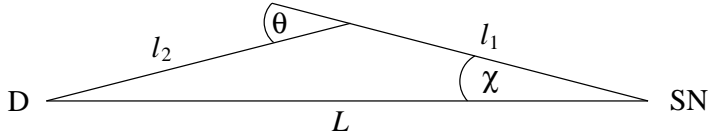}
\end{center}
\caption{\label{Geometry} The geometry of a beam from the supernova SN to
the detector D. The direct baseline has the length $L$, the baseline with one
intermediate decay has the length $l_1+l_2$, where the neutrino travels $l_1$
before decay and $l_2$ after decay.} \end{figure}

Since decaying neutrinos, which are initially traveling into directions other
than that of the detector, can produce neutrinos which are re-directed towards
the detector, we average for the calculation of $\Phi^{D,1}_{\alpha
\beta}$ over all possible decay positions at which the secondary
neutrinos may still arrive at the detector. Figure \ref{Geometry}
defines the geometry of the problem, as well as the geometrical terms in the
figure caption.
It is safe to assume that the detector is small compared with the distance
to the decay position $\sqrt{D} \ll l_2$. Thus, we can use radial
symmetry, introduce spherical coordinates, and reduce the spatial
averaging to $\int d \cos \chi/2$. For $\Phi^{D,0}_{\alpha \beta}$ we
only need to find the fraction of $\Phi_{\alpha}^{\mathrm{tot}}$ that can hit
the detector by geometry, which is $\Delta (\cos \chi)_D/2$, \ie, the cosine
range under which the detector is seen by the source. 

Since a secondary neutrino produced by decay will arrive at
the detector only with a certain probability due to kinematics, we introduce
the function $\eta_{ij}(L,l_1,l_2,D)$ describing the fraction of decay products
$j$ of parent neutrinos $i$ decaying at the position determined by $l_1$,
$l_2$, and $L$, which will still arrive at the detector. It is given by 
\begin{equation}  \eta_{ij}(L,l_1,l_2,D) \equiv
\eta_{ij}(L,\chi,l_1,D) \equiv {1 \over \Gamma_{ij}}
\int\limits_{\mathcal{D}} \left| {d \Gamma_{ij} \over d \cos \theta} \right|
{1 \over 2 \pi} d \cos \theta d \phi,  \label{defeta} \end{equation}
where $\mathcal{D}$ denotes the surface of detection and $D$ its area, which
is to be integrated over, and $\Gamma_{ij}$ is the total
decay rate. For the description of the kinematics of mass eigenstates we will
use the square root of this function, because amplitudes behave like square
roots of particles \cite{Lindner:2001fx}. We readily see from
\fig~\ref{Geometry} that $l_2(L,\chi,l_1)$ is determined by
\begin{equation} 
l_2^2 = L^2 + l_1^2 - 2 L l_1 \cos \chi, \label{l2def} \end{equation}
\ie, we may, for example, choose $l_1$ and $l_2$ or $\chi$ and $l_1$ as
sets of independent parameters. For Majoron models the redirection by decay
is because of kinematics limited by a maximum angle
$\theta_{\mathrm{max}}^{ij}$, where $\nu_i$ is the parent neutrino and $\nu_j$
the decay product. This angle can be shown to be very
small for $\Delta m_{ij}^2$ not larger than $m_i^2$ or $m_j^2$ by many orders
of magnitude and relativistic neutrinos, which is reasonable at
least for active neutrinos with no extremely hierarchical mass
spectrum. Generalizing this to any considered decay model, using $\chi$ and
$l_1$ as independent parameters, and assuming the detection area $D$ to be
small compared with the distance $l_2$, \ie, $\sqrt{D} \ll l_2$, we
determine the parameter ranges from geometry for small angles to be 
\begin{eqnarray}   \label{chirange}  \chi
& \in & [ 0, \, \theta_{\mathrm{max}}^{ij}], \\    l_1 & \in & [ 0, \, 
l_{1,ij}^{\mathrm{ max}} ], \quad \mathrm{where} \  l_{1,ij}^{\mathrm{ max}}
\equiv L \left( 1 - {\chi \over \theta_{\mathrm{max}}^{ij}} \right). 
\label{l1range}
\end{eqnarray} 
Furthermore, we have to take care of time dispersion due to different
traveling lengths. In general, $\Phi^{\mathrm{tot}}_{\alpha \beta}(t)
\propto \Phi^{\mathrm{tot}}_{\alpha}(t-t_1^{(i)}-t_2^{(j)})$, where $t_k^{(a)}
\simeq l_k [1+m_a^2/(2E^2)]$ for relativistic mass eigenstates with mass
$m_a$, $\nu_i$ is the mass eigenstate before decay, and $\nu_j$ is the one
after decay. Thus, in order to integrate the time dependence of the source
flux, we incorporate it at the time when the peaks of the wave
packets arriving at the detector were emitted at the supernova. Traveling times
are related to mass eigenstates, which are, at least within the coherence
length, to be summed over coherently. However, the supernova produces flavor
eigenstates. Thus, it will become useful to introduce the notion
of an {\em amplitude flux} $\sqrt{\Phi^{\mathrm{tot}}}$, describing the flux of
states instead of particles. 

\subsection{Operators, transition amplitudes, and fluxes}
\label{Subsec:operators}

We will now define the relevant operators, transition amplitudes, and fluxes
based on the discussions in the last section as well as in \Ref
\cite{Lindner:2001fx}. The decay rate is defined as $\alpha_{ij} \equiv
\frac{m_{i}}{\tau^{0}_{ij}}$ for $i \rightarrow j$ decay, where
$\tau^{0}_{ij}$ is the rest frame lifetime for that decay channel.
Time dilation by the factor $\gamma_i=E_i/m_i$ implies that \begin{equation}
\Gamma_{ij}^{\mathrm{observer}} = \Gamma_{ij}^{0} \gamma_i^{-1} =
{1 \over \tau_{ij}^0} \frac{m_i}{E_i} =
{\alpha_{ij} \over E_i}.
\end{equation}
The rate $\alpha_{i} \equiv \sum_{j} \alpha_{ij}$ is the overall decay
rate of the state $i$, which means that $B_{ij} \equiv
\frac{\alpha_{ij}}{\alpha_{i}}$ is the branching ratio for $i \rightarrow j$
decay. The energy $E_i$ of the mass eigenstate $i$ will be approximated by
the mean energy $E$ of the produced flavor eigenstate if the energy
corrections are of higher than first order. The operators will be defined in
terms of effective creation $\hat{a}_{i}^{\dagger}$ and annihilation
$\hat{a}_{i}$ operators operating on mass eigenstates
\cite{Lindner:2001fx}.

\begin{definition}[Disappearance operator]
$\mathcal{D}_{-}$ is the transition operator which is also often called ``decay
operator''. Effectively, it returns the amplitude for an undecayed state $i$
remaining undecayed after traveling a distance $l$ along the baseline $L$:
\begin{equation}
 \label{DMinus}
 \mathcal{D}_{-}(l) = \sum\limits_i \exp \left( - \frac{\alpha_{i} t^{(i)} }{2
E_{i}} \right) \hat{a}_{i}^{\dagger} \hat{a}_{i} \quad with \
t^{(i)} \simeq \left(1 + {m_i^2 \over 2 E^2} \right) l.
\end{equation}
\end{definition}

\begin{definition}[Appearance operator]
$\mathcal{D}_{+}$ is the differential transition operator, which destroys an
{\bf in} state and creates an {\bf out} state in $[t_1,t_1+dt_1]$ along the
baseline $L$, \ie, a new state ``appears'':
\begin{equation}
 \label{DPlus}
 \mathcal{D}_{+}(L,l_1,l_2,D) = \underset{i \neq j}{\sum\limits_i
\sum\limits_j} \sqrt{\frac{\alpha_{ij}}{E_{i}}} \ \sqrt{1+ {m_i^2 \over 2 E^2}
} \ \sqrt{ \eta_{ij}(L,l_1,l_2,D) } \, \hat{a}_{j}^{\dagger} \hat{a}_{i}.
\end{equation}
\end{definition}
The probability density, which is the square of the amplitude, has to be
integrated over $l_1$. Since $t^{(i)}  \simeq ( 1 + \frac{m_i^2}{2 E^2} ) l$
for a mass eigenstate $\nu_i$, the transition amplitude, which is  defined per
time unit, needs to be transformed into an amplitude per length unit by the
square root of this factor. Note that for a stationary problem the geometric
function $\eta$ does not depend on traveling times.

\begin{definition}[Propagation operator]
$\mathcal{E}^k$ is the operator propagating a state $\nu_k$ a distance $l$
along the baseline $L$:
\begin{equation} 
\label{Evol}
\mathcal{E}^k(l) = \exp \left( - i E_{k} l \right) 
\hat{a}_{k}^{\dagger} \hat{a}_{k}.
\end{equation}
\end{definition}
The index $k$ will need to be summed over in the calculation of the
flux, since time delays of different mass eigenstates will enter the
(macroscopic) flux formula.

\begin{definition}[Decoherence operator]
$\mathcal{S}^{I}(l,T)$ is the operator describing the loss of wave packet
coherence of a state by traveling a distance $l$ along the baseline $L$:
 \begin{equation}
 \label{SDecoh}
\mathcal{S}^{I}(l,T) = \sum_i N^{I,i} \exp \left[ - \frac{(l-v_i
 T)^2}{4 (\sigma^{I}_x)^2} \right] \, \hat{a}_i^{\dagger} a_i.
\end{equation}
\end{definition}
It turns out that this operator
has the required properties leading to the factors postulated in \Ref
\cite{Giunti:1998wq} as well as in \Sec \ref{Subsec:coherence} in
order to describe the loss of coherence for long baselines. The variable
$T$ describes the wave packet distribution in time. It will be integrated
over in the transition probability, since it is not measurable by the
target process. In addition, $\sigma^{I}_x$ is the wave packet width of the
composition $I$ of two processes (production, decay, or detection).
Furthermore, $N^{I,i}$ is a normalization factor with
\begin{equation} 
\int N^{I,i} N^{I,j} \exp \left[ - \frac{(l-v_i T)^2 + (l-v_j T)^2}{4
(\sigma^{I}_x)^2} \right] dT = \exp \left[ - \left( \frac{l}{L^{\mathrm{
coh},I}_{ij}} \right)^2  \right]. \end{equation}
Here the factor $\exp \{ - (l-v_a
T)^2/[4 (\sigma^{I}_x)^2] \}$ comes from the expansion of a sharply peaked
wave packet, which can be written as $\Psi_a(L,T) = \exp[ -i E_a T + i p_a L -
(L-v_a T)^2/(4 \sigma_x^2)]$. In the definition of the propagation
operator we ignored the factor $\exp( i p_a L)$, since it would only
give rise to common phase factors as well as an additional, already mentioned 
factor, which is negligible for $\sigma_x \ll L^{\mathrm{osc}}$.

\begin{definition}[Calculation of transition amplitudes]
\label{def:transprob}
The transition amplitude $A(\nu_{\alpha} \rightarrow
\nu_{\beta})^n=A_{\alpha \beta}^n$ by {\bf exactly} $n$ intermediate decays is
for $n=0$
\begin{equation}
 \label{adecay0}
 A_{\alpha \beta}^{0,i} = \langle \nu_{\beta} | \mathcal{E}^i(L)
\mathcal{D}_{-}(L) \mathcal{S}^{PD}(L,T)| \nu_{\alpha} \rangle
\end{equation}
and for $n=1$
\begin{equation}
 A_{\alpha \beta}^{1,ij}  = 
\langle \nu_{\beta} | \mathcal{E}^j(l_2) \mathcal{D}_{-}(l_2) 
\mathcal{S}^{XD}(l_2,T) \mathcal{D}_{+}(L,l_1,l_2,D) \mathcal{E}^i(l_1)
\mathcal{D}_{-}(l_1) \mathcal{S}^{PX}(l_1,U) | \nu_{\alpha} \rangle.
\label{adecay}
\end{equation}
The time variables $T$ and $U$ are to be
integrated over in the transition probabilities, \ie, after taking the squares
of the amplitudes. The indices $i$ and $j$, denoting the intermediate traveling
mass eigenstates, are to be summed over in the calculation of the flux. Here
$(\sigma_x^{AB})^2 \equiv (\sigma_x^A)^2 + (\sigma_x^B)^2$, where the label
$P$ refers to the production process, $D$ to the detection process, and $X$ to
the decay process. \end{definition}
Note that the order of $\mathcal{D}_{-}$, $\mathcal{S}$, and $\mathcal{E}$ is
arbitrary, since it can be shown that these operators all commute.

\begin{definition}[Calculation of fluxes]
\label{def:flux}
The supernova neutrino fluxes $\Phi^{D,0}_{\alpha \beta}(t)$ (no intermediate
decays) and $\Phi^{D,1}_{\alpha \beta}(t)$ (one intermediate decay) at the
detector D are calculated as
\begin{eqnarray}
 \label{flux0}
 \Phi^{D,0}_{\alpha \beta}(t) & = & \left| \sum\limits_i \sqrt{\Phi^{\mathrm{
tot}}_{\alpha} (t-T^{(i)})} \, A_{\alpha
\beta}^{0,i} \right|^2 {\Delta ( \cos \chi )_D \over 2}, \\
\Phi^{D,1}_{\alpha \beta}(t) & = &  \int\limits_{\cos
\hat{\theta}_{\mathrm{max}}}^1 \int\limits_{l_1=0}^{\hat{l}_1^{\mathrm{max}}}
\left| \sum\limits_i \sum\limits_j  \sqrt{\Phi^{\mathrm{tot}}_{\alpha}
(t-t_1^{(i)}-t_2^{(j)})} \, A_{\alpha \beta}^{1,ij} \right|^2  dl_1 {d \cos
\chi  \over 2}  \label{flux1}
 \end{eqnarray}
with $T^{(i)} \simeq \left(1 + {m_i^2 \over 2 E^2} \right) L$, $t_k^{(i)}
\simeq \left(1 + {m_i^2 \over 2 E^2} \right) l_k$,
$\hat{\theta}_{\mathrm{max}} = \max \theta_{\mathrm{max}}^{ij}$, and
$\hat{l}_1^{\mathrm{max}} = \max l_{1,{ij}}^{\mathrm{max}}$. \end{definition}
In $\Phi^{D,1}_{\alpha \beta}$, we must integrate over the maximal
possible range determined by the maximum of all $\theta_{\max}^{ij}$ and
$l_{1,ij}^{\mathrm{max}}$, in order to cover the appropriate spatial region.
If $(\chi,l_1)$ is out of range, which may occur in some integrands, the
function $\eta_{ij}$ in $\mathcal{D}_{+}$ will evaluate to zero.

\subsection{Limiting cases}

We now discuss some limiting cases. Most of them involve
coherence limits and can also be combined with the limiting case in \Sec
\ref{Subsubsec:neglectpaths}, \ie, neglecting different traveling
path lengths.

\subsubsection{Coherent wave packets}
\label{Subsubsec:coherent}

Let us assume that the wave packets are at all times coherent,
\ie, $L \ll L^{\mathrm{coh}}$ for all coherence lengths involved in the
problem. Hence, the decoherence operators $\mathcal{S}^{P}(l,T)$ will only give
rise to factors of unity after the integration over time $T$, and
hence, can be neglected. In addition, this implies that $l_k \simeq t_k$, \ie,
we do not have to take into account differences between traveling times and
propagation distances of different mass eigenstates. Finally, we assume that
$\theta_{\mathrm{max}}^{ij} \simeq \theta_{\mathrm{max}}$ and $l^{\mathrm{
max}}_{1,ij} \simeq l^{\mathrm{max}}_{1}$ to be independent of the indices $i$
and $j$, \ie, within the coherence length the mass eigenstates take
approximately the same traveling paths.  Therefore, we can pull the
amplitude fluxes out of the summations over the intermediate traveling mass
eigenstates, and redefine the propagation operators by absorbing the
summations. In addition, we introduce probabilities instead of amplitudes. We
then obtain the following expressions for the operators:  \begin{eqnarray} 
\mathcal{D}_{-}(l) & = & \sum\limits_i \exp \left( - \frac{\alpha_{i} l}{2 
E_{i}} \right) \hat{a}_{i}^{\dagger} \hat{a}_{i}, \\
\mathcal{D}_{+}(L,l_1,l_2,D) & = & \underset{i \neq j}{\sum\limits_i
\sum\limits_j} \sqrt{\frac{\alpha_{ij}}{E_{i}}} \ \sqrt{
\eta_{ij}(L,l_1,l_2,D) } \, \hat{a}_{j}^{\dagger} \hat{a}_{i}, \\
\mathcal{E}(l) & = & \sum\limits_i \exp \left( - i E_{i} l \right) 
\hat{a}_{i}^{\dagger} \hat{a}_{i}. \end{eqnarray} For the transition
probabilities we have
\begin{eqnarray}
 P_{\alpha \beta}^{0} & = & \left| \left< \nu_{\beta} | \mathcal{E}(L)
\mathcal{D}_{-}(L) | \nu_{\alpha} \right> \right|^2, \\
{dP_{\alpha \beta}^{1}  \over  dl_1}  & = &
\left| \left< \nu_{\beta} | \mathcal{E}(l_2) \mathcal{D}_{-}(l_2) 
\mathcal{D}_{+}(L,l_1,l_2,D) \mathcal{E}(l_1)
\mathcal{D}_{-}(l_1) | \nu_{\alpha} \right>
\right|^2,
\end{eqnarray}
and for the fluxes we find
\begin{eqnarray}
 \Phi^{D,0}_{\alpha \beta}(t) & = & \Phi^{\mathrm{tot}}_{\alpha}
(t-L) {\Delta ( \cos \chi )_D \over 2} P_{\alpha \beta}^{0}, \\
 \Phi^{D,1}_{\alpha \beta}(t) & = & \int\limits_{\cos \theta_{\mathrm{max}}}^1
\int\limits_{l_1=0}^{l_1^{\mathrm{max}}} \Phi^{\mathrm{tot}}_{\alpha}
(t-l_1-l_2) {dP_{\alpha \beta}^{1}  \over  dl_1} dl_1 {d \cos \chi  \over 2}.
\end{eqnarray}
Therefore, the expressions found in this limit are very similar to the ones in
\Ref \cite{Lindner:2001fx}, but adopted to point sources.

\subsubsection{Incoherent wave packets}
\label{Subsubsec:incoherent}

This limit corresponds to the first case mentioned in \Sec \ref{Sec:decoh},
\ie, loss of coherence between any two processes in the problem. Thus, we
assume that $L, \, l_1, \, l_2 \gg L^{\mathrm{coh}}$ for all coherence
lengths in the problem. Hence, the operators $\mathcal{S}^{I}(l,T)$ will give
rise to factors $\exp [ - ( l/L^{\mathrm{coh},I}_{ij})^2 ] \rightarrow
\delta_{ij}$ after the integration over the time $T$. It can be shown by
application of the operators that this corresponds to incoherent summation
over the intermediate traveling states, \ie, $| \sum_{ij} f_{ij} |^2 \equiv
\sum_{ij} | f_{ij} |^2$ in this limit. Therefore, we can square the amplitude
fluxes, put the summations in front of the integrations in $\Phi^{D,1}_{\alpha
\beta}$, and contract the integration limits back to the appropriate regions.
We then may define transition probabilities  \begin{eqnarray} \label{pdecay02}
 P_{\alpha \beta}^{0,i} & = & \left| \left< \nu_{\beta} | \nu_i
\right> \right|^2 \left| \left< \nu_i | \mathcal{D}_{-}(L) | \nu_i \right>
\right|^2 \left| \left< \nu_i | \nu_{\alpha} \right> \right|^2, \\
{dP_{\alpha \beta}^{1,ij}  \over  dl_1} & = &
\left| \left< \nu_{\beta} | \nu_j \right>
\right|^2 \left| \left< \nu_j | \mathcal{D}_{-}(l_2)
\mathcal{D}_{+}(L,l_1,l_2) \mathcal{D}_{-}(l_1) | \nu_i \right> \right|^2
\left| \left< \nu_i | \nu_{\alpha} \right> \right|^2,
 \label{pdecay2}
\end{eqnarray}
and fluxes
\begin{eqnarray}
 \label{flux02}
 \Phi^{D,0}_{\alpha \beta}(t) & = & \sum\limits_i \Phi^{\mathrm
{tot}}_{\alpha} (t-T^{(i)}) \, P_{\alpha
\beta}^{0,i} {\Delta ( \cos \chi )_D \over 2}, \\
\Phi^{D,1}_{\alpha \beta}(t) & = &  \sum\limits_i \sum\limits_j 
\int\limits_{\cos \theta_{\mathrm{max}}^{ij}}^1
\int\limits_{l_1=0}^{l_{1,ij}^{\mathrm{max}}} \Phi^{\mathrm{tot}}_{\alpha}
(t-t_1^{(i)}-t_2^{(j)}) \, {dP_{\alpha \beta}^{1,ij} \over dl_1}  dl_1 {d \cos
\chi  \over 2}  \label{flux12}
 \end{eqnarray}
with $T^{(i)} \simeq \left(1 + {m_i^2 \over 2 E^2} \right) L$ and $t_k^{(i)}
\simeq \left(1 + {m_i^2 \over 2 E^2} \right) l_k$.
Thus, $\mathcal{S}$ splits up the formulas for the
transition amplitudes such that neutrino oscillations vanish. 

\subsubsection{Neglecting time delays by different traveling paths}
\label{Subsubsec:neglectpaths}

Time delays due to different path lengths can be neglected when they are small
compared with time delays by different masses. The fraction of the initial flux
arriving at the detector is $D/4 \pi L^2$, since the total flux through any
sphere around the point source is equal by symmetry. This is, in principle, not
changed for decays from one mass eigenstate into another, since the
overall number of neutrinos arriving at the detector is
unchanged.\footnote{Note that we only assume small changes in direction by
decay, which means that the detector geometry does not affect this conservation
law.} Nevertheless, the relative arrival times may differ for different path
lengths. The time delay due to different path lengths $\Delta t_{1,ij}$ for
mass eigenstates $\nu_i$ and $\nu_j$, propagating before and after decay,
respectively, can for small $\theta_{\mathrm{max}}^{ij}$ be approximated by
$\Delta t_{1,ij} < \Delta t_{1,ij}^{\mathrm{max}} \simeq L
(\theta_{\mathrm{max}}^{ij})^2/2$ from geometry. For this small time interval,
delays by different masses can be ignored as second order effects. In order to
have the limiting case discussed here, it needs to be much smaller than the
time delay by different masses $\Delta t_{2,ij} = \Delta m_{ij}^2 L/(2E^2)$.
Therefore, for $\theta_{\mathrm{max}}^{ij} \ll \sqrt{\Delta m_{ij}^2}/E$
effects of different path lengths can be neglected. In fact, it can be shown
that this is equivalent to $\Delta m_{ij}^2 \ll m^2_j$ in the case of Majoron
decay. However, the time delay caused by different path lengths may still be
measurable if the absolute value of $\Delta t_1$ is longer than the time
resolution of the detector.

Since we are ignoring traveling times due to different
paths and since the total number of arriving particles is not
changed by decay, we can implement this limiting case in the geometrical
function $\eta$, describing the fraction of secondary neutrinos hitting the
detector, as
\begin{equation}
 \eta(L,\chi,l_1,D) = \frac{D}{4 \pi L^2} \,
4 \delta(1- \cos \chi).
\end{equation}
Thus, the secondary neutrinos are peaked in the forward
direction and for any $x<1$ the integral $\int_x^1 \, \eta \, d \cos \chi /2 =
D/(4 \pi L^2)$ gives the required fraction independent of the
indices $i$ and $j$. Note that this definition of $\eta$ corresponds to a
forward peaked differential decay rate $d \Gamma_{ij}/d \cos \theta \propto
\delta(1- \cos \theta)$ in \eq~(\ref{defeta}). However, it is not exactly
identical due to re-direction effects. Nevertheless, a distribution of the
differential decay rate, which is sharply peaked into the forward direction,
gives an effective $\tilde{\theta}_{\mathrm{max}}^{ij} \ll
\theta_{\mathrm{max}}^{ij}$. Then, for the case of
$\tilde{\theta}_{\mathrm{max}}^{ij} \ll \sqrt{\Delta m_{ij}^2}/E$, we have
this limit again. We will introduce in \Sec \ref{Sec:Majoron} an
$\eta$-function which corresponds to this case.

In order to incorporate the new $\eta$-function in this limiting case, we
only need to re-define one operator:
\begin{equation}
 \label{DPlusp}
 \mathcal{D}_{+} = \underset{i \neq j}{\sum\limits_i
\sum\limits_j} \sqrt{\frac{\alpha_{ij}}{E_{i}}} \ \sqrt{1+ {m_i^2 \over 2 E^2}
}  \, \hat{a}_{j}^{\dagger} \hat{a}_{i},
\end{equation}
In addition, we need to take into account that $l_2 = L-l_1$ in the
evaluation of the $\chi$-integration over the $\delta$-distribution in
the $\eta$-function. Finally, we obtain for the fluxes for $l_1 \equiv l$
\begin{eqnarray}
 \label{flux0p}
 \Phi^{D,0}_{\alpha \beta}(t) & = & \left| \sum\limits_i \sqrt{\Phi^{\mathrm{
tot}}_{\alpha} (t-T^{(i)})} \, A_{\alpha
\beta}^{0,i} \right|^2 \, {D \over 4 \pi L^2}, \\
\Phi^{D,1}_{\alpha \beta}(t) & = &  
\int\limits_{l=0}^{L} \left| \sum\limits_i \sum\limits_j \sqrt{\Phi^{\mathrm{tot}}_{\alpha}
(t-t_1^{(i)}-t_2^{(j)})} \, A_{\alpha \beta}^{1,ij} \right|^2
 dl_1 \, {D \over 4 \pi L^2}
 \label{flux1p}
 \end{eqnarray}
with $T^{(i)} \simeq \left(1 + {m_i^2 \over 2 E^2} \right) L$, $t_1^{(i)}
\simeq \left(1 + {m_i^2 \over 2 E^2} \right) l$, and $t_2^{(j)} \simeq \left(1
+ {m_j^2 \over 2 E^2} \right) (L-l)$. Simultaneous application of the
incoherent limit yields in addition formulas adjusted to the
last limiting case in a straightforward way.

\subsubsection{Decay before loss of coherence}
\label{Subsubsec:earlydecays}

In this limit, we treat the second case mentioned in \Sec \ref{Sec:decoh},
\ie,  decay rates large enough such that all of the neutrinos decay before
coherence is lost. If we assume that $L \gg
L^{\mathrm{coh}}$, the wave packets of undecayed states will loose coherence
before detection, \ie, $\Phi^{D,0}_{\alpha \beta}$ can be calculated as in
\Sec \ref{Subsubsec:incoherent}. For very early decays, \ie, $\alpha \gg
E/L^{\mathrm{coh}}$, the distance $l_2$ is much longer than the corresponding
coherence length, whereas $l_1$ is shorter. In this case, all initial
neutrinos have decayed before $l_1$ approaches the coherence
length, which also means that $\Phi^{D,0}_{\alpha \beta} \simeq 0$.
Furthermore, since $L \gg L^{\mathrm{ coh}}$, the decay products will loose
coherence before detection. In order to understand what may
happen in this limiting case, we assume interference between
different decay channels, \ie, simultaneous couplings to the decay products,
such as by Majoron coupling constants $g_{ik}>0$ and $g_{jk}>0$ for different
indices $i$, $j$, and $k$. Then, the relative phase of the states in the
incoming superposition depends on the decay position. In addition, the arrival
time depends on the decay position, because the mass eigenstates travel with
different velocities before and after decay. We expect to observe the most
interesting interference effects for $\Delta m^2 \simeq \alpha$. For fast
oscillations, \ie, $\Delta m^2 \gg \alpha$, interference effects will be
washed out by averaging over the different decay positions. For slow
oscillations, \ie, $\Delta m^2 \ll \alpha$, all particles will still have the
initial relative phase at the decay position, eliminating any
oscillating effect.

In order to calculate the transition flux $\Phi^{D,1}_{\alpha \beta}$, we
combine the formulas in \Secs \ref{Subsubsec:coherent} (before decay)
and \ref{Subsubsec:incoherent} (after decay). We define
\begin{equation}
 \bar{t}_1 \equiv \left( 1 + \frac{\overline{m^2}}{2 E^2} \right) l_1 \quad
\mathrm{with} \quad \overline{m^2} \equiv \frac{1}{N} \sum\limits_{i=1}^{N}
m_i^2,
 \label{tmean}
\end{equation}
where the mass square average is to be taken over the $N$ mass
eigenstates oscillating before decay. Since this mean traveling time will only
enter in the total source flux and $l_1$ is assumed to be quite short, this
is certainly a reasonable approximation for the source flux not changing too
much on timescales of the order $\Delta m^2 l_1/(2E^2)$.
In addition, we assume $\theta_{\mathrm{ max}}^{ij} \simeq
\theta_{\mathrm{max}}^{j}$ and $l_{1,ij}^{\mathrm{max}} \simeq 
l_{1,j}^{\mathrm{max}}$ to depend only on the decay products (the incoming
wave packets are coherent). Eventually, this yields 
\begin{eqnarray}
\label{pdecayd}  {dP_{\alpha \beta}^{1,j} \over  dl_1}  & = & \underbrace{
\Big| \langle \nu_{\beta} | \nu_j \rangle \Big|^2 \Big| \langle \nu_j |
\mathcal{D}_{-}(l_2) }_{\mathrm{incoherent}} \,  \underbrace{
\mathcal{D}_{+}(L,l_1,l_2) \mathcal{E}(l_1) \mathcal{D}_{-}(l_1) |
\nu_{\alpha} \rangle \Big|^2 }_{\mathrm{coherent}}, \\ \Phi^{D,1}_{\alpha
\beta}(t) & = &  \sum\limits_j  \int\limits_{\cos \theta_{\mathrm{max}}^{j}}^1
\int\limits_{l_1=0}^{l_{1,j}^{\mathrm{max}}} \Phi^{\mathrm{tot}}_{\alpha}
(t-\bar{t}_1-t_2^{(j)}) \, {dP_{\alpha \beta}^{1,j} \over dl_1}  dl_1 {d \cos
\chi  \over 2},
\label{flux1d}
\end{eqnarray}
with the operators from the coherent part as given in \Sec
\ref{Subsubsec:coherent}.

\section{Incoherent mass dispersion}
\label{Sec:massdisp}

In this section, we will demonstrate for the limiting case of incoherent
propagation how mass dispersion can be caused by decay even if we ignore
different traveling path lengths.\footnote{A similar effect was discussed in
\Ref \cite{Raffelt} for the arrival times of photons as decay products of
radiative neutrino decay.} This is done by simultaneous application of the
limits in \Secs \ref{Subsubsec:incoherent} and \ref{Subsubsec:neglectpaths}.
We will show that it is possible to observe a time dispersion by decay
even for a pulsed source flux. Thus, we
assume that $\Phi^{\mathrm{ tot}}_{\alpha}(t) = N_{\alpha} \, \delta(t+L)$
which describes a neutrino pulsed produced at $t=-L$, where $N_{\alpha}$ is
the total number of neutrinos of flavor $\nu_{\alpha}$ emitted by the
supernova. In this case, it will be possible to detect
massless neutrinos at $t=0$ and massive neutrinos delayed by $\Delta t^{(i)} =
{m_i^2 l \over 2 E^2}$. Decay leads to a time dispersion, since a mass
eigenstates travels with a different velocity before and after decay.
Its arrival time depends therefore on the decay position. 

Combing the limiting cases in \Secs \ref{Subsubsec:incoherent} and
\ref{Subsubsec:neglectpaths} as well as applying the transition probabilities
and operators, we obtain for $\Phi_{\alpha \beta}^{D,1}$
\begin{eqnarray}
 \Phi^{D,1}_{\alpha \beta}(t) & = & \underset{i \neq j}{\sum\limits_i
\sum\limits_j} \int\limits_{l=0}^{L} \frac{N_{\alpha} D}{4 \pi L^2}
\delta \left( t- \left(1+{m_i^2 \over 2 E^2} \right) l - \left(1+{m_j^2 \over
2 E^2} \right) (L-l) +L \right) \nonumber \\
& \times & | U_{\alpha i}|^2 | U_{\beta j}|^2 
e^{-{\alpha_j \over E} (L-l) \left( 1+ {m_j^2 \over 2 E^2} \right) }
{\alpha_{ij} \over E} e^{-{\alpha_i \over E} l \left( 1+ {m_i^2 \over 2 E^2}
\right) }
\left(1 + {m_i^2 \over 2 E^2} \right) dl.
\end{eqnarray}
The $\delta$-distribution in this formula, describing the neutrino pulse,
implies that
\begin{equation}
 t = {\Delta m_{ij}^2 \over 2 E^2} l + {m_j^2 \over 2 E^2} L
\end{equation}
or, for $m_i \neq m_j$,
\begin{equation}
 l = {2 E^2 \over \Delta m_{ij}^2} \left( t - {m_j^2 \over 2
E^2} L \right).
 \label{leq}
\end{equation}
The condition $0 \le l \le L$, for which the $\delta$-distribution evaluates to
a non-zero value by the integration, is then equivalent to
\begin{equation}
 {m_j^2 \over 2 E^2} L \le t_{ij} \le {m_i^2 \over 2 E^2} L.
 \label{tinterval}
\end{equation}
Thus, the signal arrives at the detector between the signals of
the undecayed heavy mass eigenstate and the light one produced by the
intermediate decay. Note that the allowed time interval depends on the indices
$i$ and $j$. In addition, we need to take into account that  $\delta \left(
f(l) \right) = \delta (l-a) / |f'(a) |_{f(a)=0} = 2 E^2/\Delta m_{ij}^2 \delta
(l-a)$ with $a$ equal to $l$ as given in \eq~(\ref{leq}). Finally,
integrating over $l$ leads to
\begin{eqnarray}
 \Phi^{D,1}_{\alpha \beta}(t) & = & \underset{i \neq j}{\sum\limits_i
\sum\limits_j} \frac{N_{\alpha} D}{4 \pi L^2}
| U_{\alpha i}|^2 | U_{\beta j}|^2 {2 E \alpha_{ij} \over \Delta m_{ij}^2}
\left( 1 + {m_i^2 \over 2 E^2} \right) e^{ -{\alpha_j \over E} L
\left( 1+ {m_j^2 \over 2 E^2} \right) } \nonumber \\
 & \times & \exp \left\{ \left[ \alpha_j  \left( 1+ {m_j^2 \over 2
E^2} \right) - \alpha_i  \left( 1+ {m_i^2 \over 2 E^2} \right) \right] {2 E
\over \Delta m_{ij}^2} \left( t - {m_j^2 \over 2 E^2} L \right)
\right\}. \nonumber\\
\end{eqnarray}
In the limit of stable decay products, \ie, $\alpha_j \rightarrow 0$, and
by ignoring corrections of the order $m^2/E^2$ to the signal height, this
reduces to
\begin{equation}
 \Phi^{D,1}_{\alpha \beta}(t) = \underset{i \neq j}{\sum\limits_i
\sum\limits_j} \frac{N_{\alpha} D}{4 \pi L^2}
| U_{\alpha i}|^2 | U_{\beta j}|^2 {2 E \alpha_{ij} \over \Delta m_{ij}^2}
\exp \left[  - \alpha_i  {2 E \over \Delta m_{ij}^2}
\left( t - {m_j^2 \over 2 E^2} L \right) \right].
 \label{dispstable}
\end{equation}
Thus, the flux pulse of the source is smeared
out to an exponentially dropping signal at the detector. Here the allowed
range for $t_{ij}$ depends on the mass eigenstates $i$ and $j$ and is given in
\eq~(\ref{tinterval}). One can immediately see that the smearing is determined
by the coefficient $\alpha_i E/\Delta m_{ij}^2$ in the exponential. We obtain
maximal smearing for small $\alpha_i E/\Delta m_{ij}^2$, \ie, small
$\alpha_i$'s or large $\Delta m_{ij}^2$'s. A numerical analysis for typical
values of $\Delta m_{ij}^2$, $E$, and $L$ shows that the $\alpha_{ij}$'s, and
the Majoron coupling constants $g_{ij}$'s, respectively, can be far below any
currently assumed upper limit for observing exponential dropping of this
function (for constraints on the $g_{ij}$'s, see \Refs
\cite{Barger:1982vd,Acker:1992ej,Kachelriess:2000qc,Tomas:2001}). However, if
$\alpha_i$ is too small, the factor $\alpha_{ij}$ will suppress the term
$\Phi^{D,1}_{\alpha \beta}$ completely.

\begin{figure}[ht!]
\begin{center}
\includegraphics*[height=12cm,angle=270]{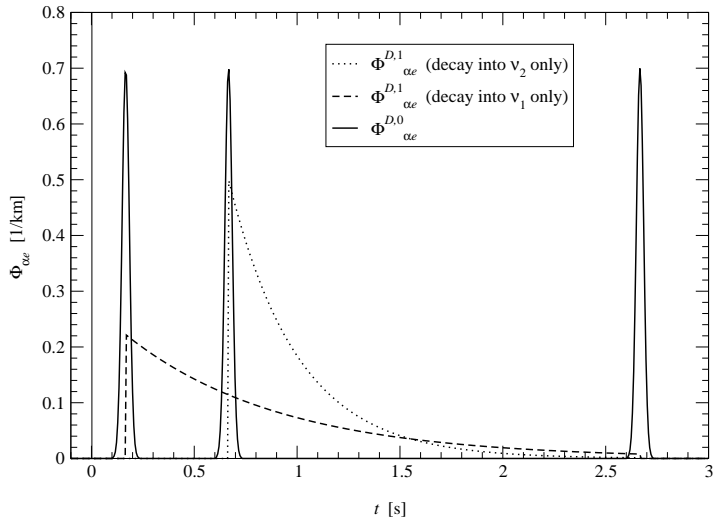}
\end{center}
\caption{\label{Dispersion} The (separated) signals of the
decay products $\Phi^{D,1}_{\alpha e}$ and
those of the undecayed particles $\Phi^{D,0}_{\alpha e}$ for the scenario
constructed in the text. The
$\delta$-distributions in $\Phi^{D,0}_{\alpha e}$ are plotted as Gaussian
signals, where the leftmost refers to the lightest and the rightmost to the
heaviest mass eigenstate. The vertical line at $t=0$ indicates the reference
time when massless neutrinos would arrive.} \end{figure}

\begin{figure}[ht!]
\begin{center}
\includegraphics*[height=12cm,angle=270]{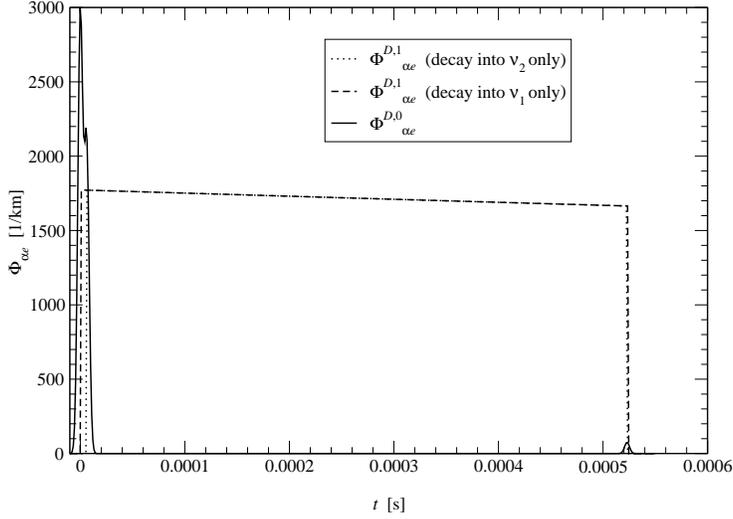}
\end{center}
\caption{\label{Dispersion2} The signals of the decay products
$\Phi^{D,1}_{\alpha e}$ and those of the undecayed particles
$\Phi^{D,0}_{\alpha e}$ for the scenario with the parameter values
$\Delta m_{32}^2 = 2.5 \cdot 10^{-3} \, \mathrm{eV}^2$, $\Delta
m_{21}^2 = 3.65 \cdot 10^{-5} \, \mathrm{eV}^2$, $m_1 = 10^{-5} \,
\mathrm{eV}$ (absolute neutrino mass value needed here), $\theta_{12} =
\theta_{23} = 45^\circ$ (bimaximal mixing), $\theta_{13} = 5^\circ$, and
$\delta_{CP} = 0$. The other parameter values are the same as for
Fig.~\ref{Dispersion}. Note that $\alpha$ can be equal to $\mu$ or
$\tau$. For $\alpha = e$ the result will be different, since the
mixing is not trimaximal anymore as in Fig.~\ref{Dispersion}. Furthermore,
note that the scale of the fluxes is not applicable to the $\delta$-fluxes
illustrated as peaked signals, but the heights of these peaks reflect
their relative values.}
\end{figure}

We illustrate the effect for the following three-neutrino scenario:
Maximal mixing, \ie, $| \nu_{\alpha} \rangle = {1 \over \sqrt{3}} | \nu_1
\rangle +{1 \over \sqrt{3}}  | \nu_2 \rangle+ {1 \over \sqrt{3}} | \nu_3
\rangle$ ($\theta_{12} = \theta_{13} = \theta_{23} = 45^\circ$ and
$\delta_{CP} = 0$), decay of $\nu_3$ into $\nu_2$ or $\nu_1$ only,
\ie, $\alpha_{32} = \alpha_{31} = 10^{-21} \, \mathrm{GeV \,
km^{-1}}$, $E=10 \, \mathrm{MeV}$, $L=10^{22} \, \mathrm{m} \simeq 32
\, \mathrm{kpc}$, $N_{\alpha} = 9 \cdot 10^{5} \, (4 \pi L^2)/D$,
$m_3=4 \, \mathrm{eV}$, $m_2=2 \, \mathrm{eV}$, and $m_1=1 \,
\mathrm{eV}$, which means that $\Delta m_{32}^2 = 12 \, \mathrm{eV}^2$
and $\Delta m_{21}^2 = 3 \, \mathrm{eV}^2$. We can use
\eqs~(\ref{flux0}) and (\ref{dispstable}) to 
evaluate $\Phi^{D,0}_{\alpha e}$ and $\Phi^{D,1}_{\alpha e}$, where the sums
are split up in order to see the different signals from different mass
eigenstates. Figure~\ref{Dispersion} shows the separated signals of the decay
products $\Phi^{D,1}_{\alpha e}$ as well as the ones from the undecayed
particles $\Phi^{D,0}_{\alpha e}$.
It has been observed \cite{Beacom:1999em} that the SN1987A data can be
reasonably fit to a decaying exponential with time constant $\tau \simeq 3
\, \mathrm{s}$. Figure~\ref{Dispersion} indicates
that we can easily find parameter sets in order to have an effect with a
similar time dependence.
However, for the currently assumed parameter values for active
neutrinos, as shown in \fig~\ref{Dispersion2}, the timescale of this
effect would be rather short.
In addition, the two different decay channels could, mainly due to the small
$\Delta m_{21}^2$, not be resolved anymore. Thus, in order to be able
to identify all relevant individual effects in the plots, we will further on
choose parameter values which amplify these effects more clearly, but which
are not the values currently favored by data. There are basically two
arguments supporting this point of view. First, the best-fit
parameters may change when making global fits with neutrino decay
included, and therefore, the mass squared differences may become
larger. Second, more recently, other models being relevant for
this sort of discussion have been introduced, such as in \Ref~\cite{Ma:2001ip},
in which the decay of a light sterile neutrino into active neutrinos was
proposed. In this case, the relevant mass squared differences could
again be rather large. Such light sterile neutrinos could be produced by
oscillated supernova neutrinos close to or within the supernova.

\section{Dynamics of Majoron decay}
\label{Sec:Majoron}

\begin{figure}[ht!]
\begin{center}
\includegraphics*[height=9cm]{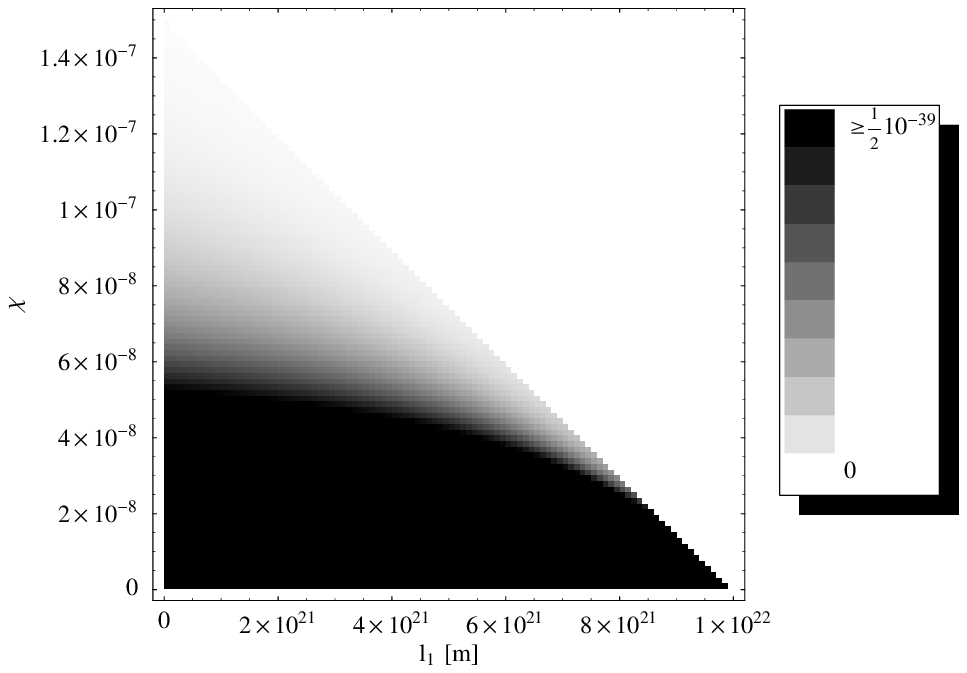}
\end{center}
\caption{\label{etapseudo} Density plot with a linear shading scale of the
function $\eta_{ij}(\chi,l_1)$ for pseudoscalar Yukawa couplings in the
Lagrangian (decay into neutrinos). Here $D=10000 \, \mathrm{m}^2$,
$L=10^{22} \, \mathrm{m} \simeq 32 \, \mathrm{kpc}$, $E_i=E=10 \,
\mathrm{MeV}$, $m_i=2 \, \mathrm{eV}$, and $m_j=1 \, \mathrm{eV}$. For the
geometry of the problem, see \fig~\ref{Geometry}.} \end{figure}

\begin{figure}[ht!] \begin{center} \includegraphics*[height=9cm]{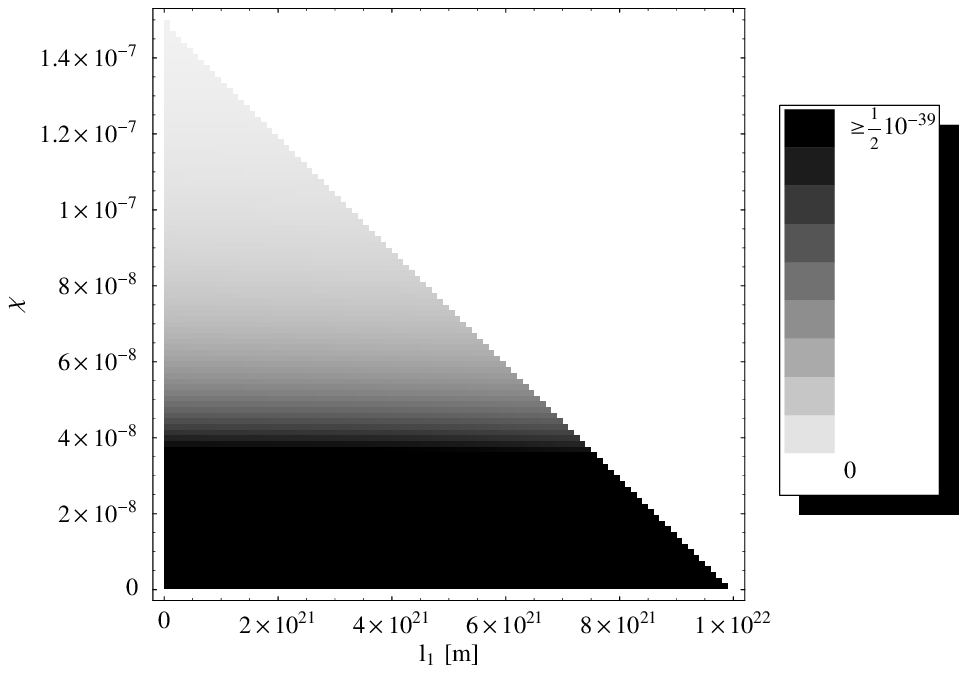}
\end{center}
\caption{\label{etascalar} Density plot with a linear shading scale of the
function $\eta_{ij}(\chi,l_1)$ for scalar Yukawa couplings in the Lagrangian
(decay into neutrinos). Here $D=10000 \, \mathrm{m}^2$, $L=10^{22} \,
\mathrm{m} \simeq 32 \, \mathrm{kpc}$, $E_i=E=10 \, \mathrm{MeV}$, $m_i=2 \,
\mathrm{eV}$, and $m_j=1 \, \mathrm{eV}$. For the
geometry of the problem, see \fig~\ref{Geometry}. } \end{figure}

\begin{figure}[ht!]
\begin{center}
\includegraphics*[height=9cm]{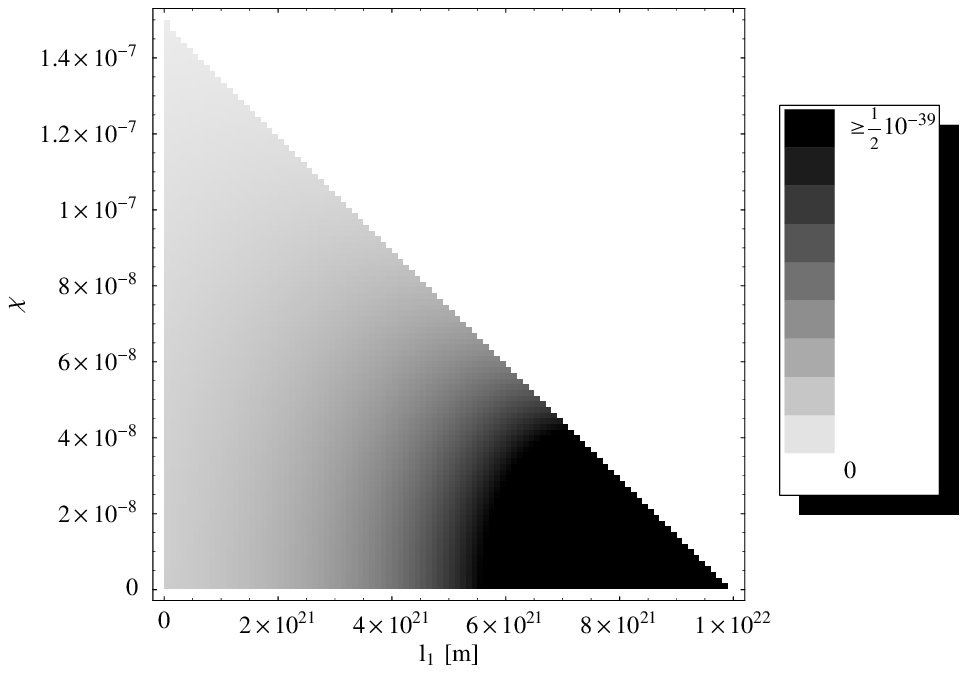}
\end{center}
\caption{\label{etaanti} Density plot with a linear shading scale of the
function $\eta_{ij}(\chi,l_1)$ for pseudoscalar or scalar Yukawa couplings in
the Lagrangian (decay into antineutrinos). Here $D=10000 \, \mathrm{m}^2$,
$L=10^{22} \, \mathrm{m} \simeq 32 \, \mathrm{kpc}$, $E_i=E=10 \,
\mathrm{MeV}$, $m_i=2 \, \mathrm{eV}$, and $m_j=1 \, \mathrm{eV}$. For the
geometry of the problem, see \fig~\ref{Geometry}. }
\end{figure}

Let us now give an approximation for the function $\eta_{ij}(L,\chi,l_1,D)$
defined in \eq~(\ref{defeta}) for Majoron decay of
relativistic neutrinos. It describes the fraction of decay products arriving
at the detector for decay at $(\chi,l_1)$ (\cf, \fig~\ref{Geometry}).
Since only a small fraction of neutrinos will decay close to the detector, we
can assume that
\begin{equation}  
 L \ge l_2 \gg \sqrt{D}/\theta_{\mathrm{max}}^{ij},
\end{equation}
where $\theta_{\mathrm{max}}^{ij}$ is given by \cite{Lindner:2001fx}
\begin{equation}
 \theta_{\mathrm{max}}^{ij} = {m_j \over 2 E_i} \left( x^2-1 \right), \qquad x
\equiv {m_i \over m_j}.
 \label{thetamaxmaj}
\end{equation}
Here the index $i$ refers to the parent neutrino and $j$ to the decay
product. Hence, we can treat the detector as a point target
and approximate \eq~(\ref{defeta}) by
\begin{equation}
 \eta_{ij}(L,\chi,l_1,D) = {1 \over
\Gamma_{ij}} \left| {d \Gamma_{ij} \over d \cos \theta}
\right|_{(\cos \theta)_D} {1 \over 2 \pi} \Delta \Omega = {1 \over
\Gamma_{ij}} \left| {d \Gamma_{ij} \over d \cos \theta} \right|_{(\cos
\theta)_D} {D \over 2 \pi l_2^2}
 \label{eta1}
\end{equation}
with $l_2$ given by \eq~(\ref{l2def}). From
\begin{equation}
 E_i-E_j=| \mathbf{p}_i - \mathbf{p}_j| = \sqrt{ |
 \mathbf{p}_i |^2+ |\mathbf{p}_j |^2 - 2 | \mathbf{p}_i | |\mathbf{p}_j | \cos
\theta},
 \label{KinMain}
\end{equation}
which determines the kinematics of the process, we find for
small $\theta$ two energies $E_j$ corresponding to one angle $\theta$
\begin{equation}
 E_j^{\pm}(\theta) = \frac{E_i (1+x^2)}{2 \left( {E_i^2 \theta^2 \over
m_j^2} + x^2 \right)} \left[ 1 \pm \sqrt{1-4 \frac{{E_i^2 \theta^2 \over
m_j^2}+x^2}{(1+x^2)^2} } \right].
\end{equation}
Equation~(\ref{eta1}) can be written as
\begin{equation}
\eta_{ij}(L,\chi,l_1,D) = {1 \over \Gamma_{ij}} {D \over 2 \pi l_2^2}
\left\{  \left| {d \Gamma_{ij} \over d E_j} \right|_{E_j^+} \left| {d E_j
\over d \cos \theta} \right|_{E_j^+} + \left| {d \Gamma_{ij} \over d E_j}
\right|_{E_j^-} \left| {d E_j \over d \cos \theta} \right|_{E_j^-} \right\}.
 \label{eta2}
\end{equation}
The differential and total decay rates in this equation are for (pseudoscalar
or scalar) Yukawa couplings in the Lagrangian for decay into neutrinos or
antineutrinos given in \Refs \cite{Lindner:2001fx,Kim:1990km},
respectively, for the Lagrangians introduced there. From \eq~(\ref{KinMain})
we read off  \begin{equation}  \left| {d E_j \over d \cos \theta}
\right|_{E_j^{\pm}} = \frac{|\mathbf{p}_i| |\mathbf{p}_j|}{| E_i - E_j|}
\simeq \frac{E_i}{\left| {E_i \over E_j}-1 \right|}.
 \label{dEdct}
\end{equation}
Since we use $\chi$ and $l_1$ as independent parameters, we need to
express $\theta$ in terms of $\chi$. To first approximation for small angles,
which is reasonable for a small $\theta_{\mathrm{max}}^{ij}$, we find from
geometry \begin{equation}
 \theta \simeq  \left( 1+ {l_1 \over l_2} \right) \chi \simeq \left( 1 +
\frac{l_1}{L-l_1} \right) \chi.
 \label{thetachi}
\end{equation}
Now $\eta$ can be evaluated numerically and is plotted in
\figs~\ref{etapseudo}, \ref{etascalar}, and \ref{etaanti} for
(pseudoscalar or scalar) Yukawa couplings in the Lagrangian and decay into
neutrinos and antineutrinos. 
 Note that the parameter
ranges determined in \eqs~(\ref{chirange}) and (\ref{l1range}) imply
that $\eta$ evaluates to zero, \ie, nothing arrives at the detector if the
parameter set lies within the upper-right triangle over the diagonal. In this
case, the secondary neutrino cannot be re-directed towards the detector
anymore for kinematical reasons. The plot for antineutrinos shows, in
comparison to the neutrino plots, that the direct path ($\chi=0$) is
suppressed for small $l_1$, because of a spin flip. For Yukawa scalar
couplings $\eta$ turns out to be quite independent of the angle $\chi$. In all
cases, we can expect heavy suppression for too large $\chi$. 

The magnitude of $\eta$ is mainly determined by the
$D/l_2^2$-dependence of \eq~(\ref{eta1}) (antineutrinos) or by the factor $d
E_j/d \cos \theta$ in \eq~(\ref{eta2}) (neutrinos). In our example, $D/l_2^2
= \mathcal{O}(D/L^2) = 10^{-40}$ for large $l_2$. One can show that $d
E_j/d \cos \theta$ diverges for $E_j=E_j^+$ at $\theta=0$, which evaluates
there to $E_j^+=E_i$. Thus, in the limit $L \ge l_2 \gg
\sqrt{D}/\theta_{\mathrm{max}}$, the $\eta$-function is difficult to evaluate
numerically in the transition probabilities. Nevertheless, since the integral
over the differential cross section over the whole parameter range has to give
$\Gamma_{ij}$, the mean differential decay rate is of the order
$\Gamma_{ij}/(\theta_{\mathrm{max}}^{ij})^2$, and thus, $\bar{\eta} =
\mathcal{O}\left( D/(l_2 \theta_{\mathrm{max}}^{ij})^2 \right) \gg D/l_2^2$.
Hence, for neutrinos the main contribution to the differential decay rate
comes from the divergent, forward peaking part, and we may, in this case,
neglect time delays by different path lengths as it was done in
\Sec \ref{Sec:massdisp}. For antineutrinos the forward direction is suppressed
by the spin flip and becomes finite, which means that we may approximate
the differential cross section by its mean value in order to obtain from
\eq~(\ref{eta1})
\begin{equation}  
 \bar{\eta}_{ij}(L,\chi,l_1,D) \equiv  {1
\over \Gamma_{ij}} \underbrace{\overline{ \left( {d \Gamma_{ij} \over d \cos
\theta} \right)}}_{\Gamma_{ij} \over (\tilde{\theta}_{\mathrm{max}}^{ij})^2/2}
{1 \over 2 \pi} \Delta \Omega= {D \over \pi \left[ l_2 \left( L,l_1,\chi
\right) \right]^2} {1 \over \left( \tilde{\theta}_{\mathrm{max}}^{ij}
\right)^2}. \label{etamean} 
\end{equation}
Here $\tilde{\theta}_{\mathrm{max}}^{ij} \le
\theta_{\mathrm{max}}^{ij}$, where is an effective maximum angle. For
neutrinos, \figs~\ref{etapseudo} and \ref{etascalar} indicate that
$\tilde{\theta}_{\mathrm{max}}^{ij} \ll \theta_{\mathrm{max}}^{ij}$ could
also be used as an approximation. For antineutrinos,
\fig~\ref{etaanti} is dominated by the $1/l_2^2$-dependence, which means that
we may use the approximation $\tilde{\theta}_{\mathrm{ max}}^{ij} =
\theta_{\mathrm{max}}^{ij}$. Note that for
$\tilde{\theta}_{\mathrm{max}}^{ij} < \theta_{\mathrm{max}}^{ij}$ we also have
to change $\theta_{\mathrm{max}}^{ij} \rightarrow
\tilde{\theta}_{\mathrm{max}}^{ij}$ and $l_{1,ij}^{\mathrm{max}} \rightarrow
\tilde{l}_{1,ij}^{\mathrm{max}} =  L ( 1- \chi/ \tilde{\theta}_{\mathrm{
max}}^{ij} )$ in the integration limits in \eq~(\ref{flux1}).

\section{Fluxes for incoherent propagation}
\label{Sec:general}

We will now provide the most general expressions for the
fluxes in the limit of incoherent wave packet propagation introduced in
\Sec \ref{Subsubsec:incoherent}, where we also take into account time delays
by different path lengths. Again we assume a neutrino pulse produced at
the supernova, \ie,  $\Phi^{\mathrm{tot}}_{\alpha}(t) =
N_{\alpha} \delta(t+L)$, such that massless neutrinos would arrive at the
detector at time $t=0$. For other sources the results for this
$\delta$-flux can be used to describe a time-dependent source. We neglect
corrections to the signal heights coming from $l \neq t$, since these are of
the order $m^2/E^2$. However, we are interested in time dispersion induced by
non-zero masses. Noticing that $2 \pi \Delta ( \cos \chi )_D L^2 = D$ for the
direct baseline, we obtain from \eqs~(\ref{flux0}) and (\ref{pdecay02})
\begin{equation}  \Phi_{\alpha \beta}^{D,0}(t)   \simeq \sum\limits_i 
N_{\alpha} \delta \left( t - {m_i^2 \over 2 E^2} L \right) {D \over 4 \pi L^2}
 | U_{\alpha i} |^2 | U_{\beta i} |^2 e^{- {\alpha_i \over E} L}.
\end{equation} {}From \eqs~(\ref{flux1}) and (\ref{pdecay2}) we find
\begin{eqnarray} 
 \Phi^{D,1}_{\alpha \beta}(t) & \simeq & \underset{i \neq j}{\sum\limits_{i}
\sum\limits_{j}} \int\limits_{\cos \theta_{\mathrm{max}}^{ij}}^1
\int\limits_{l_1=0}^{l_{1,ij}^{\mathrm{max}}} N_{\alpha} \delta \left( t+L- \left(
1+ {m_i^2 \over 2 E^2} \right) l_1 - \left( 1+ {m_j^2 \over 2 E^2} \right) l_2
\right) \nonumber \\ 
& & \times \underbrace{| U_{\alpha i} |^2 | U_{\beta j} |^2 e^{- {\alpha_j
\over E} l_2} {\alpha_{ij} \over E} \eta_{ij}(L,\chi,l_1,D) e^{- {\alpha_i
\over E} l_1}}_{{dP_{\alpha \beta}^{1,ij}  \over  dt^{(i)}_1}}  dl_1 {d \cos
\chi  \over 2}.
\end{eqnarray}
Integrating over $l_1$, we must observe that
the limits of the $l_1$-integration depend on $\chi$ and we have to take
into account that the $\delta$-distribution evaluates to a non-zero
value only for $0 \le l_1 \le l_{1,ij}^{\mathrm{max}} = L ( 1- \chi/
\theta_{\mathrm{max}}^{ij} )$. This can often be done by adjusting the
integration limits of the $\chi$-integration, but here it will lead to quite
complicated expressions for $l_1(t,\chi)$ and $\chi(t,l_1)$, respectively.
We therefore introduce the function
\begin{equation}
 \xi_{ij}(t,\chi) \equiv \left\{ \begin{array}{ll} 1 \quad & \mathrm{for} \ \ 0
\le l_1(t,\chi) \le l_{1,ij}^{\mathrm{max}} = L \left( 1- {\chi \over
\theta_{\mathrm{max}}^{ij}} \right) \\ 0 \quad & \mathrm{otherwise}
\end{array} \right.
 \label{xidef}
\end{equation}
to describe the region where the integrand contributes.
Furthermore, we have to take into account the transformation of the
$\delta$-distribution $\delta \left( f(l)
\right) = \delta (l-a) / |f'(a) |_{f(a)=0}$ with $a=l_1(t,\chi)$ being the
solution of $f(a)=0$ (see below). In this case, we can
approximate $|f'(a)|_{f(a)=0}$ by
\begin{equation}
  \mu(t,\chi) \equiv |f'(a)|_{f(a)=0} \simeq {\Delta m_{ij}^2 \over 2 E^2} + {L
\over l_2 \left(l_1(t,\chi),\chi \right)} {\chi^2 \over 2}.
\end{equation}
Thus, for small $\chi$ we arrive at
\begin{eqnarray} \Phi^{D,1}_{\alpha \beta}(t) &
\simeq & \underset{i \neq j}{\sum\limits_{i} \sum\limits_{j}} N_{\alpha} 
{\alpha_{ij} \over 2E} | U_{\alpha i} |^2 | U_{\beta j} |^2 \nonumber \\
 & & \qquad \times \int\limits_{0}^{\theta_{\mathrm{max}}^{ij}} e^{- {1 \over E}
(\alpha_i l_1 + \alpha_j l_2)} \, \eta_{ij}(L,\chi,l_1,D) \, {\xi_{ij}(t,\chi)
\over \mu(t,\chi)} \, \chi \, d \chi
 \label{phix}
\end{eqnarray}
with
\begin{equation}
 l_1 = \frac{t- \left( 1+ {m_j^2 \over 2 E^2} \right) l_2 + L}{1+
{m_i^2 \over 2 E^2}}
 \label{l1implicit} 
\end{equation}
coming from the $\delta$-distribution. In addition, \eq~(\ref{l2def}) implies
that  $l_2^2 = L^2 + l_1^2 - 2 L l_1 \cos \chi$.
This leads together with \eq~(\ref{l1implicit}) to a quadratic
equation in $l_1(t,\chi)$ or $l_2(t,\chi)$. Analysis of $l_1(t,\chi)$ shows
that we obtain a unique, quite lengthy solution for $l_1>0$, which we
will not present here. It can also be shown that $t(\chi,l_1)$ grows
monotonously with growing $\chi$ or $l_1$, which implies that the
earliest arrival time is $Lm_j^2/2E^2$. The latest arrival time can be
obtained from the maximum of $t(\chi,l_1^{\mathrm{max}}(\chi))$, which is
again a quite lengthy expression. Since we neglect
corrections of the signal height of the order $m^2/E^2$ and second order
corrections, we can approximate \eq~(\ref{l1implicit}) in the exponential of
\eq~(\ref{phix}) by \begin{equation}  l_2 \simeq t - l_1 + L. 
\end{equation}
We finally obtain
\begin{eqnarray}
 \Phi^{D,1}_{\alpha \beta}(t) &
\simeq & \underset{i \neq j}{\sum\limits_{i} \sum\limits_{j}} N_{\alpha} 
{\alpha_{ij} \over 2E} | U_{\alpha i} |^2 | U_{\beta j} |^2 
 \int\limits_{0}^{\theta_{\mathrm{max}}^{ij}} e^{- {1 \over E} \left(
\alpha_i l_1(t,\chi) + \alpha_j (L-l_1(t,\chi)) \right)} \nonumber \\
 & & \qquad \times  \eta_{ij}(L,\chi,l_1(t,\chi),D) \, {\xi_{ij}(t,\chi)
\over \mu(t,\chi)} \, \chi \, d \chi. 
\label{phifinal}
\end{eqnarray}

For an arbitrary time-dependent source
$\tilde{\Phi}^{\mathrm{tot}}_{\alpha}(t) = N_{\alpha} f(t+L)$ with
$\int_{-\infty}^{\infty} f(t+L) dt = 1$ we can calculate the flux at the
detector $\tilde{\Phi}^{D}_{\alpha \beta}$ from the above expressions for
$\Phi^{D,0}_{\alpha \beta}$ and $\Phi^{D,1}_{\alpha \beta}$ by
\begin{equation} \tilde{\Phi}^{D}_{\alpha \beta}(t) =
\int\limits_{-\infty}^{\infty} \Phi^{D}_{\alpha \beta}(t-t') f(t'+L) dt' = {1
\over N_{\alpha}} \int\limits_{-\infty}^{\infty} \Phi^{D}_{\alpha \beta}(t-t')
\tilde{\Phi}^{\mathrm{tot}}_{\alpha}(t'+L) dt'. \end{equation} 
Similarly, one can fold these
fluxes with energy dependencies and detector properties.

\begin{figure}[ht!]
\begin{center}
\includegraphics*[height=9cm]{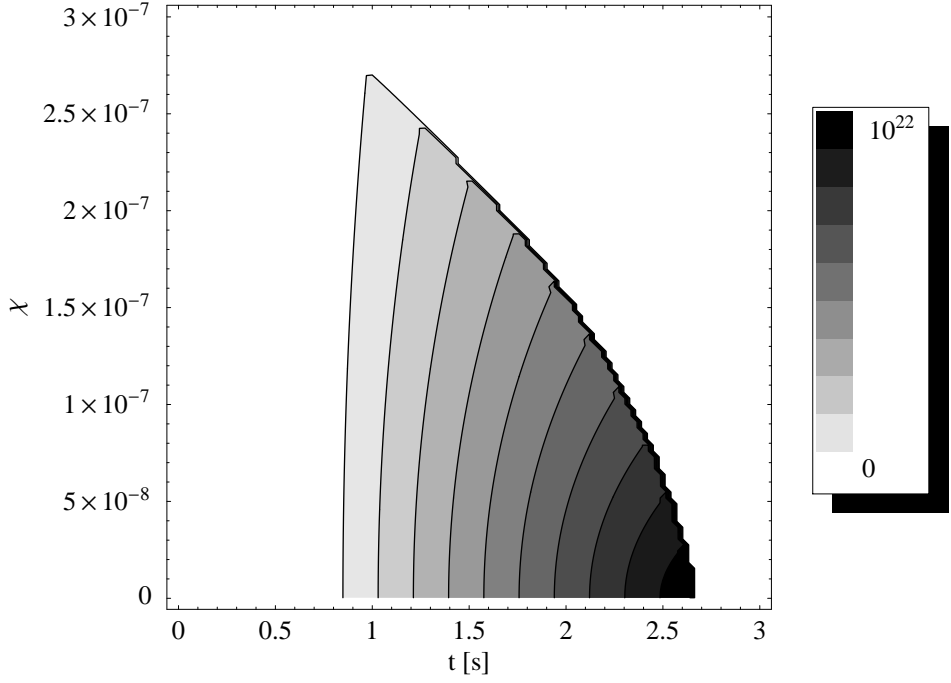}
\end{center}
\caption{\label{phidef1} The linearly scaled contour lines of the function
$\xi(t,\chi) \, l_1$ for
$\tilde{\theta}_{\mathrm{max}}=\theta_{\mathrm{max}}$.} \end{figure}

\begin{figure}[ht!]
\begin{center}
\includegraphics*[height=9cm]{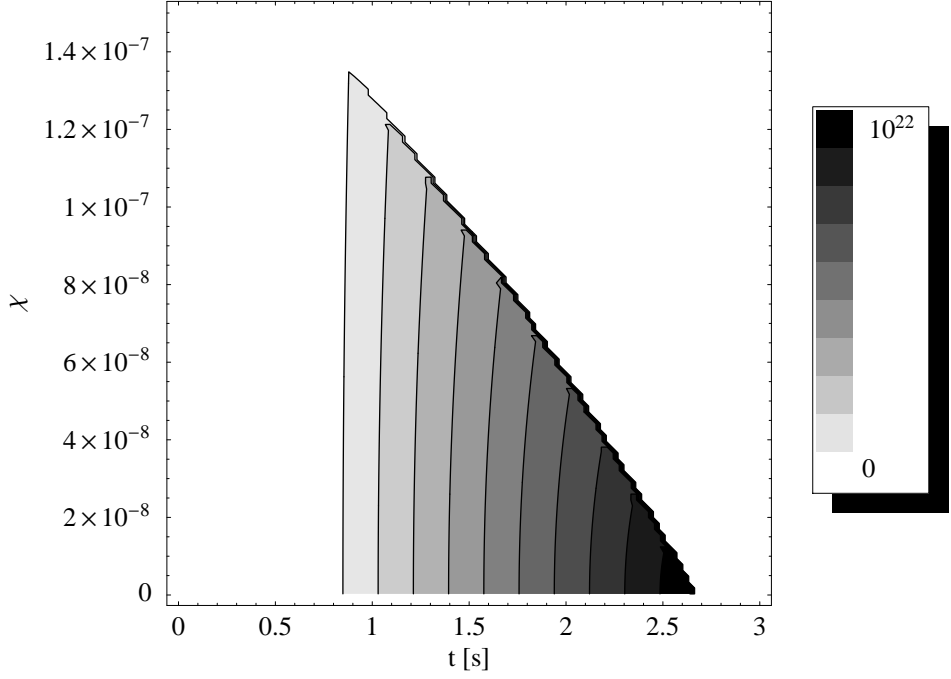}
\end{center}
\caption{\label{phidef2} The linearly scaled contour lines of the function
$\xi(t,\chi) \, l_1$ for
$\tilde{\theta}_{\mathrm{max}}=\theta_{\mathrm{max}}/2$.} \end{figure}

Let us now illustrate the effects of decay in \eq~(\ref{phifinal}) by an
example similar to the one at the end of \Sec \ref{Sec:massdisp}. We are using
the same parameters except from $\alpha_{31}=0$ and
$\alpha_{32}=c \, \alpha_0 = c \, 10^{-21} \, \mathrm{GeV} \,
\mathrm{km}^{-1}$, where $c = \mathrm{const.}$, since one decay channel is
sufficient for showing the effects. For $\eta$ we use the approximation
$\bar{\eta}$ in \eq~(\ref{etamean}), which also means that the upper
integration limit in \eq~(\ref{phifinal}) is to be replaced by
$\tilde{\theta}_{\mathrm{max}}^{32} \equiv \tilde{\theta}_{\mathrm{max}} <
\theta_{\mathrm{max}}^{32} \equiv \theta_{\mathrm{max}}$, as well as
$\theta_{\mathrm{max}}^{32}$ in \eq~(\ref{xidef}), which  represents the
integration limits for $l_1$. It turns out that the numerical evaluation of
\eq~(\ref{phifinal}) is quite sensitive to the approximations as
well as to the parameter sets, which means that the solutions only
can be used to demonstrate the qualitative behavior. From \eq~(\ref{xidef}) we
know that the function $\xi(t,\chi)$ can only take the values $0$ or
$1$, giving the areas where the integrand in \eq~(\ref{phifinal}) is defined.
Figures \ref{phidef1} and \ref{phidef2} show these areas times the traveling
path length before decay $l_1(t,\chi)$ for $\tilde{\theta}_{\mathrm{max}} =
\theta_{\mathrm{max}}$ and $\tilde{\theta}_{\mathrm{max}} =
\theta_{\mathrm{max}}/2$.
Note that the $\chi$-scales are different in these two 
plots. The white regions indicate $\xi=0$, \ie, the integrand in
\eq~(\ref{phifinal}) does not contribute. One can see that
the contours of equal $l_1$  become quite independent of $\chi$ for small
$\tilde{\theta}_{\mathrm{max}}$, \ie, the arrival time $t$ becomes dominated
by different velocities due to different masses (\cf, \fig~\ref{phidef1}). For
large $\tilde{\theta}_{\mathrm{max}}$ we can see the effects of the different
path lengths, curving the contours towards large $t$, \ie, late arrival, for
large $\chi$ (\cf, \fig~\ref{phidef2}). Thus, in comparison to the example in
\Sec \ref{Sec:massdisp}, we expect enhancements for late time arrivals and
suppressions for early time arrivals, because of the delays due to
different traveling paths. For
the case of Majoron decay in \Sec \ref{Sec:Majoron} we demonstrated that for
neutrinos the function $\eta(\chi,l_1)$ is favoring small $\chi$, which means
that the approximation $\tilde{\theta}_{\mathrm{max}} \ll
\theta_{\mathrm{max}}$ should be good. In this case, the problem reduces
to the one in \Sec \ref{Sec:massdisp}. For antineutrinos, the approximation
$\tilde{\theta}_{\mathrm{max}} = \theta_{\mathrm{max}}$ is better. 

\begin{figure}[ht!]
\begin{center}
\includegraphics*[height=12cm,angle=270]{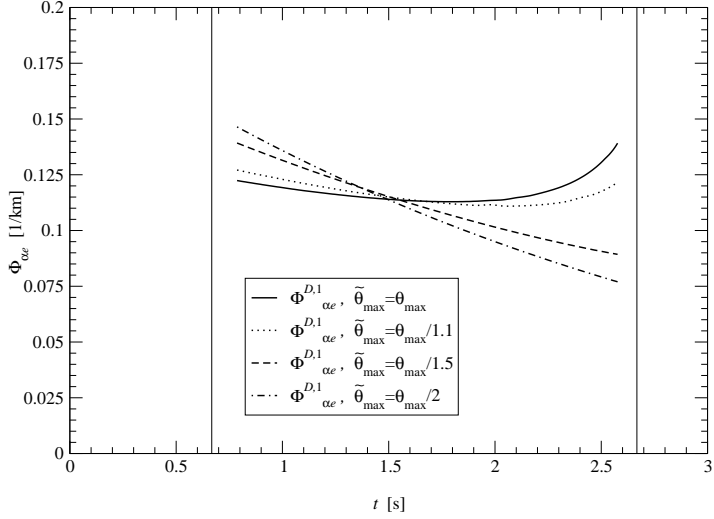}
\end{center}
\caption{\label{phitheta} The flux of decay products $\Phi^{D,1}_{\alpha
e}$ for the decay rate $\alpha=\alpha_0=E/L$ and different values of
$\tilde{\theta}_{\mathrm{ max}}$.} \end{figure}

Figure \ref{phitheta} shows the qualitative behavior of $\Phi^{D,1}_{\alpha e}$
for different $\tilde{\theta}_{\mathrm{max}}$.
The vertical lines\footnote{The functions are not plotted in the whole
possible range, because close to the limits the numerical evaluation becomes
quite unstable, since we need to integrate an almost divergent function over an
infinitesimally small interval.} indicate the
arrival times of the light and heavy mass eigenstates on the direct
paths, described by the $\delta$-distributions in $\Phi^{D,0}_{\alpha e}$,
whereas massless particles would arrive at $t=0$. One can see that the smaller
$\tilde{\theta}_{\mathrm{max}}$, the more exponential the time dependence
becomes, such as in \Sec \ref{Sec:massdisp} for forward traveling only. For
large $\tilde{\theta}_{\mathrm{max}}$ late time arrivals
are preferred, corresponding to particles delayed by longer path lengths.
In principle, some particles could even arrive after the heavy mass
eigenstate traveling on the direct path, but these had to decay
quite late in order to keep the slow velocity of the heavy mass eigenstate as
long as possible. Geometry ($\theta<\theta_{\mathrm{max}}$) implies, however,
that we have the shortest overall traveling paths for $l_1$ close to $0$ or
$L$. This means that this is only possible for the case where path
length effects dominate the mass dispersion, \ie,
$\tilde{\theta}_{\mathrm{max}}^{ij} \gg \sqrt{\Delta m_{ij}^2}/E$, which we do
not consider in this example.

\begin{figure}[ht!]
\begin{center}
\includegraphics*[height=12cm,angle=270]{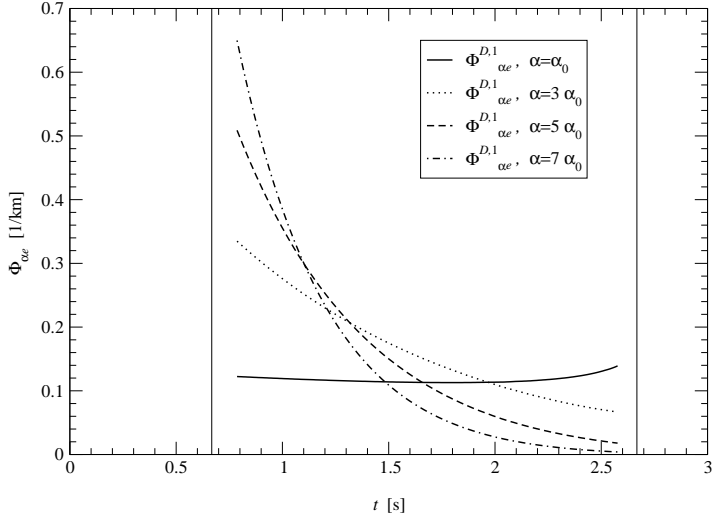}
\end{center}
\caption{\label{phialpha} The flux of decay products $\Phi^{D,1}_{\alpha
e}$ for $\tilde{\theta}_{\mathrm{max}}=\theta_{\mathrm{max}}$ and different
values of decay rates $\alpha$ as multiples of $\alpha_0=E/L$.} \end{figure}

In \fig~\ref{phialpha}, the effect of different decay rates on
$\Phi^{D,1}_{\alpha e}$ is illustrated.
Note that $\alpha_0$ is chosen such
that $\alpha_0 L/E =1$. For larger decay rates we see a behavior closer to the
one in \Sec \ref{Sec:massdisp}, \ie, exponential dropping. In this case, most
particles will decay early along the traveling path. Geometry implies in
addition that the overall path length is close to its minimum for $l_1 \simeq
0$. Therefore, for large decay rates path length effects can again be ignored.
We conclude that different traveling path lengths only have to be taken into
account in computing the supernova neutrino signal for large enough
$\tilde{\theta}_{\mathrm{max}}$ and small enough decay rates. However, since
$\Phi^{D,1}_{\alpha e}$ is directly proportional to the decay rate, too
small decay rates will make the flux $\Phi^{D,1}_{\alpha e}$ vanish.

\section{Early coherent decays}
\label{Sec:earlydecays}

As already mentioned above in \Sec
\ref{Subsubsec:earlydecays}, one can see some peculiarities for incoherent
propagation over the entire baseline $L$, \ie, $L \gg L^{\mathrm{coh}}$, but
for decay before loss of coherence, $\alpha \gg E/L^{\mathrm{coh}}$. Let us
again assume a flux pulse $\Phi^{\mathrm{tot}}_{\alpha} = N_{\alpha}
\delta(t+L)$. Neglecting corrections to the signal height of the order
$m^2 L/(2E^2)$, as well as path length dependencies for simplicity, \ie,
combing the results of \Secs \ref{Subsubsec:neglectpaths} and
\ref{Subsubsec:earlydecays} in a straightforward way, we obtain from
\eqs~(\ref{flux1p}-\ref{flux1d})
\begin{eqnarray} 
{dP_{\alpha \beta}^{1,j} \over dl } & = & | U_{\beta j} |^2 \left|
\sum\limits_i U_{\alpha i}^* e^{- \frac{\alpha_j (L-l)}{2 E}}
\sqrt{\alpha_{ij} \over E} e^{ - i E_i l} e^{- \frac{\alpha_i l}{2 E}  }
\right|^2, \\  
\Phi^{D,1}_{\alpha \beta}(t) & = & \frac{D N_{\alpha}}{4 \pi
L^2} \sum\limits_j  \nonumber \\
 & \times & \int\limits_0^L \delta \left( t+L-l \left( 1 +
{\overline{m^2} \over 2 E^2} \right) -(L-l) \left( 1 +
{m_j^2 \over 2 E^2} \right) \right) {dP_{\alpha \beta}^{1,j} \over dl } dl,
\end{eqnarray} 
where $\overline{m^2}$ refers to the mean mass square of the heavy mass
eigenstates before decay, as introduced in \Sec
\ref{Subsubsec:earlydecays}. Defining $K_{ijkl}^{\alpha \beta} \equiv
U_{\alpha i}^* U_{\beta j} U_{\alpha k} U_{\beta l}^*$ and assuming the
secondary neutrinos are stable, \ie, $\alpha_j \equiv 0$, as well as $\Im K =
0$, we find after some algebra \begin{eqnarray}
 \Phi^{D,1}_{\alpha \beta}(t) & = & \frac{D N_{\alpha}}{4 \pi L^2}
\underset{i \neq j, \, k \neq j}{\sum_i \sum_k \sum_j} \frac{2 E
\sqrt{\alpha_{ij} \alpha_{kj}}}{\overline{m_{ik}^2}- m_j^2} K_{ijkj}^{\alpha
\beta} \nonumber \\  &\times& \exp \left[ -
 \frac{( \alpha_i + \alpha_k ) E}{\overline{m_{ik}^2}- m_j^2} \left( t -
\frac{m_j^2}{2 E^2} L \right) \right] 
\cos \left[
\frac{\Delta m_{ik}^2 E}{\overline{m_{ik}^2}- m_j^2} \left( t - \frac{m_j^2}{2
E^2} L \right) \right]. \nonumber\\
\end{eqnarray}
Here $\overline{m_{ik}^2} \equiv (m_i^2+m_k^2)/2$ is
evaluated for the heavy mass eigenstates before decay, and the condition
$m_j^2 L/(2E^2) \le t \ll \overline{m_{ik}^2} L/(2E^2)$, coming from the
integration limits, has to be satisfied. This is a very interesting result,
since we may see a time-dependent oscillation if two $\alpha_{ij}$'s are
non-zero, \ie, if we have two decay channels with only one neutrino decay
product. The time $t$ indirectly measures the decay position via the flight
time of the decay product. Thus, it also measures the relative phase of the
incoming states at this position, which induces the oscillations. However,
since all $\alpha$'s are assumed to be quite large, $\alpha \gg
E/L^{\mathrm{coh}}$, the exponential indicates that the dominant contribution
comes from $t \simeq m_j^2 L/(2E^2)$. Therefore, only a small fraction of the
possible time interval $\Delta t = (\overline{m_{ik}^2}-m_j^2)L/(2E^2)$  will
be covered by the signal. The oscillating effect will hence not be observable
in most cases. Thus, we have to integrate the observed signal over time,
leading to \begin{equation}
N_{\alpha \beta}^{D,1} \equiv
\int\limits_{\frac{L m_j^2}{2E^2}}^{\frac{\overline{m_{ik}^2} L}{2E^2}}
\Phi_{\alpha \beta}^{D,1}(t) dt =  \frac{D N_{\alpha}}{4 \pi L^2} \underset{i
\neq j, \, k \neq j}{\sum_i \sum_k \sum_j} 2 K_{ijkj}^{\alpha \beta}
\frac{\sqrt{\alpha_{ij} \alpha_{kj}} (\alpha_i + \alpha_k)}{(\alpha_i +
\alpha_k)^2 + (\Delta m_{ik}^2)^2}.
\end{equation}
In comparison to entirely
incoherent propagation without interference of decay channels, which
corresponds to $\delta_{ik}$ in the summation, we observe the following two
peculiarities:
\begin{itemize}
\item  Interference of decay channels $\alpha_{ij}
\neq 0$, $\alpha_{kj} \neq 0$, and $i \neq k$ creates additional terms in
the summation, which, in principle, enhance the number of detectable
neutrinos.  \item  The mentioned interference terms 
are reduced with increasing $\Delta m_{ik}^2$ due to phase shifts in early
decays. This reduction can be neglected for $\Delta m_{ik}^2 \ll \alpha_i +
\alpha_k$, because the decay length is in such a scenario much shorter than
the oscillation length and the neutrinos will have decayed before any
oscillations. For $\Delta m_{ik}^2 \gg \alpha_i + \alpha_k$ the interference
terms will vanish by averaging over all possible decay positions
and the corresponding phases.
\end{itemize}

\section{Summary and conclusions}
\label{Sec:summary}

In this paper, we have combined neutrino decay and neutrinos oscillations for
radially symmetric sources, such as supernovas. We have calculated
time-dependent fluxes at the detector by taking into account decoherence
due to the long baselines, interference effects within the coherence
lengths, time dependence of the source, different flight times of
different mass eigenstates, and different traveling paths for neutrinos
re-directed in decay. We observed two interesting arrival time dispersion
effects, which can in some cases be of the same order of magnitude:
\begin{enumerate}
\item
The time of arrival depends on the decay position, since for
visible neutrino decay products neutrinos travel with different velocities
before and after decay. The
decay positions are distributed in an exponential manner, which means that we
will also observe an exponential decrease of the detected flux
in time.
\item
The time of flight depends on the path. For radially symmetric sources,
neutrinos with different traveling paths arrive at the detector, since decay
may change the direction of the secondary neutrinos.
This effect suppresses early time arrivals and favors late time arrivals,
working in the opposite direction to (1). 
\end{enumerate}
We have demonstrated that the first effect can mimic an exponentially
decreasing flux at the detector, such that it can be fit to
SN1987A data, for very small decay rates. This shows that supernova
properties, which are inferred from a description without neutrino decay, can
be altered or even completely changed.
Moreover, we have shown for Majoron
decay that the second effect has only to be taken into account for
antineutrinos and not too large decay rates $\alpha L/E \simeq 1$, if the
masses of the participating neutrinos are of similar order of magnitude. For
the case of $\Delta m_{ij}^2 \gg m_j^2$ for active neutrinos, the
second effect would be much larger than the one discussed in this paper.

Coherence is not only related to the
production process, but also to the detection or decay process. Thus, for
small coupling constants in the Lagrangians coherence lengths can in some
cases be quite large. In some sense, neutrino decay may act as a ``coherence
lens'' by making the wave packets collapse. However, finite lifetime
constraints, such as \eq~(\ref{ConstraintM}) for Majoron decay, have to be
satisfied in order not to wash out interference effects. There are also some
counter-intuitive peculiarities, which may be observed for very large decay
rates. Even if only incoherent mass eigenstates (or at least
one) arrive at the detector, there may be interference effects modifying the
event rates. This is because neutrinos for very early decays may decay as
long as they are still coherently propagating. For extreme choices of the
masses and decay rates, as well as simultaneous coupling
to the decay product by two different mass eigenstates in the Lagrangian (\eg,
for Majoron coupling constants $g_{31} \simeq g_{21} \neq 0$), one can, in
principle, have a time-dependent oscillation of the signal at the detector. In
this case, the sensitivity-dependent neutrino oscillation is transferred into
a time dependence of the signal by the time of flight concept of the different
mass eigenstates. Note that this may also happen even if only one mass
eigenstate is stable, which finally arrives at the detector.

Since extremely small coupling constants in the
Lagrangian can cause neutrino decay over typical supernova distances,
neutrino decay should always be taken into account in the calculation of
transition probabilities and fluxes. The observations of
supernova neutrinos from SN1987A indicate so far only that there is at least
one stable mass eigenstate. We have shown that
a number of effects involving neutrino decay into different mass
eigenstates may alter the event rates at the detector even for only one
arriving mass eigenstate. Thus, supernova neutrinos are an excellent probe to
test neutrino decay. However, similar effects can be induced by
details of the supernova explosion, which means that a separation of
decay effects and supernova details may be difficult. 

\ack
We would like to thank A. Baha Balantekin and Walter Grimus for useful
discussions and comments.

This work was supported by the Swedish Foundation for International
Cooperation in Research and Higher Education (STINT) [T.O.], the Wenner-Gren
Foundations [T.O.], the ``Studienstiftung des deutschen Volkes'' (German
National Merit Foundation) [W.W.], and the ``Sonderforschungsbereich 375
f{\"u}r Astro-Teilchenphysik der Deutschen Forschungsgemeinschaft''.

\bibliographystyle{h-elsevier}
\bibliography{references_r}

\end{document}